\documentclass[notitlepage]{report}
\RequirePackage{amsthm,amsmath,amsfonts,amssymb}
\RequirePackage[authoryear]{natbib}
\RequirePackage[colorlinks,citecolor=blue,urlcolor=blue]{hyperref}
\RequirePackage{graphicx}
\usepackage{subfigure}
\usepackage[left=0.5in, right=0.75in, top=0.5in, bottom=1in]{geometry}

\theoremstyle{plain}

\newtheorem{theorem}{Theorem}[section]
\newtheorem{corollary}{Corollary}[section]

\theoremstyle{remark}

\title{Large Sample Inference with Dynamic Information Borrowing}
\date{\today}
\author{Sergey Tarima\\ Data Science Institute, Medical College of Wisconsin \and Silvia Calderazzo \\ Division of Biostatistics, German Cancer Research Center (DKFZ)
\and Mary Homan\\ Institute of Health and Equity, Medical College of Wisconsin}

\begin{document}
\maketitle
\begin{abstract}
Large sample behavior of dynamic information borrowing (DIB) estimators is investigated. Asymptotic properties of several DIB approaches (adaptive risk minimization, adaptive LASSO, Bayesian procedures with empirical power prior, fully Bayesian procedures, and a Bayes-frequentist compromise) are explored against shrinking to zero alternatives. As shown theoretically and with simulations, local asymptotic distributions of DIB estimators are often non-normal. A simple Gaussian setting with external information borrowing illustrates that none of the considered DIB methods outperforms others in terms of mean squared error (MSE): at different conflict values, the MSEs of DIBs are changing between the MSEs of the maximum likelihood estimators based on the current and pooled data. To uniquely determine an optimality criterion for DIB, a prior distribution on the conflict needs be either implicitly or explicitly determined using data independent considerations. Data independent assumptions on the conflict are also needed for DIB-based hypothesis testing. New families of DIB estimators parameterized by a sensitivity-to-conflict parameter “s” are suggested and their use is illustrated in an infant mortality example. The choice of “s” is determined in a data-independent manner by a cost-benefit compromise associated with the use of external data.\end{abstract}


\section{Introduction}

Borrowing information from external data (e.g., historical controls) is an attractive approach for overcoming limitations of a given (current) dataset. Information borrowing can be achieved by jointly modelling the data generating processes of the external and current datasets, which  typically relies on existence of shared parameters. 
The main concern with the use of external information is that the shared parameters may not exist, which is addressed by the dynamic information borrowing (DIB) methods.
\cite{pocock1975} outlined guidelines for `acceptability' of historical controls for use in the current trial analysis; these guidelines broadly apply to the differences between the current and external trial populations as well as how these trials were conducted. Researchers can use covariate information to reduce the differences, but typically this only partially addresses the problem. Such differences can negatively impact inference: bias and variance in estimation tasks and error rates in hypothesis testing problems. The issue of the negative impact repeatedly appears in statistical literature; see \cite{freidlin2013borrowing, jin2023bayesian, galwey2017supplementation}. In contrast to other information borrowing methods, DIB methods borrow external information depending on the observed conflict: external data are used more if no conflict is determined and suppress external data otherwise. In our manuscript we focus on DIB-based inference in the presence of the conflict of an unknown magnitude which cannot be reduced or further reduced with the use of observed covariates. 

We consider statistical inference about a population parameter $\theta = EX$ for a random variable $X$ observable in the current data, where $EX$ denotes the expected value of $X$. If $\theta$ is known to be shared between the two datasets ($\theta = EX = EY$), $Y$ is a random variable observable in the external data, then both datasets should be \textit{pooled} together for inferential purposes. If, however, there is a possibility that $EY = \beta = \theta + \delta$, where the \textit{conflict} $\delta \ne 0$, the use of external data becomes questionable. Several groups of DIB estimators are explored in our manuscript: test-then-pool (\cite{viele2014use}), mean squared error minimization (\cite{Tarima2009,yu2015adaptive}), adaptive LASSO (ALASSO), (\cite{zou2006adaptive,li2023frequentist}), 
Bayesian power prior methods (\cite{ibrahim2000,ibrahim2015}) with estimated tuning parameters (e.g., \cite{gravestock2017adaptive,calderazzo2023coping}), purely Bayes approaches with a prior on the conflict, and compromise estimators bounding limiting risk of Bayes estimators (\cite{efron1972limiting}).
Out of these methods only ALASSO enjoys the oracle property: asymptotically equivalent to the pooled estimator if $\delta=0$ and the maximum likelihood estimator (MLE) based on the current data only  if $\delta \ne 0$.

The literature on the use of external information without conflict ($\delta=0$) is large, see a review \cite{qin2022selective}. \cite{tarima2006} incorporated uncertain additional information from external observations on a different but correlated quantity. 
\cite{chen2024integrating} incorporated external summary information without a known measure of uncertainty of external information. Our focus is only on the situation when 
$\delta=0$ is under question. 

This manuscript investigates asymptotic distributions of DIB estimators and their impact on estimation and hypothesis testing in the \textit{local} asymptotic framework, where $\delta = h/\sqrt{n}$, $h\ne 0$ is a \textit{local conflict} and $n$ is the sample size of the current data. When $n$ is large, a case where $\delta$ is a fixed value different from zero is not very interesting as many DIB methods easily detect $\delta\ne 0$ and asymptotically fully suppress the impact of external data. This motivates our focus on a fixed $h$. Local asymptotics gives additional and somewhat unexpected insights into DIB behavior, as their properties have been mostly evaluated for fixed deltas. More reasons for the choice of local asymptotic framework are given in a dedicated Section \ref{sec:locas}.

To simplify explanation of asymptotic comparisons of DIB estimators, we immediately start Section \ref{sec:estimators} with the asymptotic case by considering normal data with unit variances. No covariates, no multiple external datasets, and no non-Gaussian distributions of data are needed to highlight important similarities and differences of large sample properties of DIBs.



Section \ref{sec:estimators} introduces known DIB estimators and proposes Generalized DIB (GDIB) estimators (see Section \ref{sec:expanded}). These GDIB estimators incorporate a sensitivity-to-conflict parameter used to select a DIB estimator with desired properties. Section \ref{sec:asymptotics} derives the asymptotic properties of the DIB estimators, which are typically \textit{not} normal. Section \ref{sec:simulations} compares $\sqrt{n\cdot MSE}$ (SRMSE) of DIB estimators to the SRMSE's lower bound. Section \ref{sec:hypothesis.testing} reports on hypothesis testing. Section \ref{Illustrative_Example1} presents an illustrative example, which uses a family of DIB estimators introduced in Section \ref{sec:expanded}. Illustrative example shows how additional model parameters (e.g., sensitivity to conflict) can be chosen to secure the desired asymptotic characteristics. Section \ref{sec:summary} summarizes the paper with a short discussion and key findings. 

\section{Importance of local asymptotics framework} \label{sec:locas}
The practical value of local alternatives is not sufficiently emphasized in a curriculum of a typical statistics graduate program, its applied use is rarely discussed, and local alternatives are often perceived as not directly applicable to practice. However, most statistical inference relying on the central limit theorem (CLT) is done within the local asymptotic framework.

Examples of estimators which analysis relies on CLT include but are not limited to regression coefficients of generalized linear models (e.g. logistic regression), quasi-likelihood based models, semi-parametric models (e.g. Cox regression), M- and U- statistics (e.g. Mann-Whitney U), some non-parametric estimators (e.g., kernel based density estimators). Under certain regularity conditions (see, for example, Section 5 of \cite{ferguson2017course}), $\sqrt{n}\left(\hat\theta-\theta\right) \overset{d}{\to} N(0,\sigma^2)$, where $\hat\theta$ is an estimator of $\theta$ and $0<\sigma^2<\infty$. When a researcher relies on a CLT, the effect size becomes inversely proportional to the square root of the sample size, which immediately places the whole statistical inference in a local asymptotic setting: a simple sample size formula for asymptotic testing of $H_0:\beta = \theta$ versus $H_1:\beta=\theta + \delta > \theta$ is $n = \left[z_{0.95} + z_{0.8}\right]^2\sigma^2/\delta^2 \approx 6.18 \sigma^2/\delta^2$ at $0.05$ and $0.2$ type 1 and 2 error rates. Thus, $h = \sqrt{n}\delta \approx \sqrt{6.18}\sigma^2$ and $\delta = O(n^{-0.5})$. 

What happens if asymptotics against a fixed alternative is considered? The power of the test converges to 100\% for any fixed bounded away from zero alternative. This type of asymptotics sometimes applies to epidemiological studies with large sample sizes so that many reasonable hypothesis testing tasks ends up with very low p-values. Our manuscript does not consider asymptotics against fixed alternatives.

Local alternatives are used in definitions of Pitman asymptotic relative efficiency of non-parametric tests (\cite{nikitin1995asymptotic}) and asymptotic power (\cite{ferguson2017course}). Local alternatives are related to the concept of \textit{contiguety} introduced by Le \cite{cam1960locally}. Interested readers are referred to \cite{vandervaart1998} and \cite{sidak1999theory}. Local alternatives are also important for comparing MSEs of estimators: asymptotically non-zero contributions of variance and squared bias to the MSE are only possible when the magnitude of the bias is $O(1/\sqrt{n})$.  

When a CLT meets local alternatives, large sample properties become more complex. \cite{lecam1960} showed local asymptotic normality but, in contrast to regular CLTs, both the mean and the variance of the limiting normal distribution depend on the local alternatives. In some settings, the limiting distribution is a mixture of normal distributions (\cite{kahn2009local}), in others is a mixture of truncated distributions (\cite{tarima2024cost}). 

\section{Dynamic Information Borrowing Methods} \label{sec:estimators}
Many statistical methods rely on CLTs when their large sample properties are evaluated. Thus, in our pursuit of deriving large sample properties for DIB estimators, we directly focus on Gaussian settings assuming that the sample sizes of the current and external datasets are sufficiently large. Without loss of generality, we also assume unit standard deviations (SD) for current and external data: at large samples the unknown SDs can be consistently estimated and the observations can be standardized to the unit scale. 

Let $\mathbf{X}_n = \left(X_1,\ldots,X_n\right)$ be an independent random sample (the current data) with $X_i \sim N(\theta,1)$, where $\theta$ is the parameter of interest. The MLE of $\theta$,  $\hat\theta = n^{-1}\sum_{i=1}^{n} X_i$, is unbiased $(E\hat\theta = \theta)$, and absorbs all relevant information about $\theta$ from $\mathbf{X}_n$; $f_{\hat\theta}(x\vert\theta) \overset{d}{=} N(\theta,n^{-1})$. 
Similarly, let $\mathbf{Y}_m = \left(Y_1,\ldots,Y_m\right)$ be an independent sample (an external sample) with $Y_i \sim N(\beta,1)$, $\beta = \theta + \delta$. The MLE of $\beta$ based on $\mathbf{Y}_m$ is $\hat\beta=m^{-1}\sum_{i=1}^{m} Y_i$; $f_{\hat\beta}(x\vert\beta) = N(\beta,m^{-1})$. Both samples are used to calculate the MLE of $\delta$, $\hat\delta = \hat\beta - \hat\theta$. 

If $\delta=0$, the \textit{pooled} estimator $\hat\theta^{p} = (n\hat\theta + m\hat\beta)/(n+m)$ is more precise than $\hat\theta$: $var(\hat\theta) = n^{-1}$, $var(\hat\beta) = m^{-1}$, and $var(\hat\theta^p) = (n+m)^{-1}$. If $\delta\ne0$, $\hat\theta^{p}$ becomes biased its bias is not suppressed as $n\to\infty$. DIB estimators are ``floating'' between $\hat\theta$ and $\hat\theta^p$ depending on $\hat\delta$.

\subsection{Test-then-pool [TTPool]} \label{sec:test-then-pool}
The TTPool estimator was suggested by \cite{viele2014use} and stems out from traditional hypothesis testing. The approach starts with testing $H_0:\delta = 0$ using $\hat\xi=\left(1/n+1/m\right)^{-1/2}\hat\delta \sim N(0,1)$ under $H_0$. Specifically, if $\hat\xi^2 \ge c$, at some critical value $c$, $H_0$ is rejected and $\hat\theta$ is used to estimate $\theta$. Otherwise $\hat\theta^p$ is used instead. Thus, TTPool estimator is $$\hat\theta^{ttp} = I\left(\hat\xi^2 \ge c\right) \hat\theta + I\left(\hat\xi^2 < c\right)\hat\theta^{p}.$$ Under $h=0$, $\hat\xi^2\overset{d}{=}\chi^2_1$, where $\chi^2_1$ a chi-squared random variable with $1$ degrees of freedom. A typical choice of $c$ is 0.95-level quantile of $\chi^2_1$, $c=3.84$.

\subsection{Minimum Mean Squared Error[OMMSE and AMMSE]}
\label{sec:mmse}
To borrow external information, consider 
$\theta^{\lambda} = \hat\theta + \lambda\left(\hat\beta-\hat\theta\right) = \hat\theta + \lambda\hat\delta$, where $\lambda$ defines a mixing proportion. 
Following \cite{Tarima2009} and \cite{tarima2020estimation}, the smallest MSE in the class $\theta^{\lambda}$ is reached at $ \lambda_0 = -cov\left(\hat\theta,\hat\delta\right)E^{-1}\left(\hat\delta^2\right)$ by
\begin{align*}
\hat\theta^{ommse} &= \hat\theta +\lambda_0 \hat\delta = \hat\theta + \frac{m}{n +m +nm\delta^2} \hat\delta
\end{align*}
with 
$MSE\left(\hat\theta^{ommse}\right) = n^{-1}\left(1 - m/(n + m +nm\delta^2)\right)$. Under $\delta = 0$, $\theta^{\lambda}$ becomes unbiased for all choices of $\lambda$ and $\hat\theta^{ommse}=\hat\theta^p$
with $var\left(\hat\theta^{p}\right) = 1/(n+m)$. For a fixed $n$, $var\left(\hat\theta^{p}\right)>0$ and only gets close to zero when $m$ is much larger than $n$ $(m>>n)$. The optimal MMSE (OMMSE) estimator $\hat\theta^{ommse}$ cannot be applied in practice because $\delta$ is unknown. The adaptive version of OMMSE (AMMSE) substitutes $\delta$ with $\hat\delta$:
$$\hat\theta^{ammse} = \hat\theta + \frac{m}{n + m + nm\hat\delta^2} \hat\delta.
$$

\subsection{Adaptive LASSO [ALASSO]} 
\label{sec:alasso}
Recently ALASSO (\cite{zou2006adaptive}), a modification of the LASSO regression (\cite{tibshirani1996regression}), was used for dynamic information borrowing (see \cite{li2023frequentist} and \cite{kanapka2024frequentist}). In our settings, the ALASSO's penalized log-likelihood ($PL$) 
\begin{align*}
PL(\theta,\delta\vert \hat\theta,\hat\beta) &\propto - \sum_{i=1}^n\left(X_i - \theta \right)^2 - \sum_{i=1}^m\left(Y_i - \theta - \delta\right)^2 -
(n+m)^{\tau} \frac{\vert \delta\vert }{\vert\hat\delta\vert} 
\\ & \propto - \frac{n}{n+m}\left(\hat\theta - \theta\right)^2  - \frac{m}{n+m}\left(\hat\beta - \theta -\delta\right)^2 - \frac{\vert \delta\vert }{(n+m)^{0.5-\tau}\sqrt{n+m}\vert\hat\delta\vert},
\end{align*}
where the tuning parameter $(n+m)^{\tau-1}$ and the adaptive ``weight'' $\vert\hat\delta\vert^{-1} = \vert\hat\beta-\hat\theta\vert^{-1}$ 
make the penalized loglikelihood differ from the (un-adaptive) LASSO. Per Theorem 2 in \cite{zou2006adaptive}, $\tau\in(0,0.5)$ ensures the oracle property of the ALASSO estimator  
$$\hat\theta^{alasso}(\tau) = \arg_{\theta}\max_{\theta,\delta} PL(\theta,\delta\vert \hat\theta,\hat\beta).$$

The ALASSO estimator is equivalent to the posterior mode under an (empirical) Laplace prior on $\delta$. In particular, the posterior density would be of the form
$$
\pi(\theta, \delta \vert \hat\theta, \hat\beta) \propto 
\exp \left\{-\frac{n}{2}(\hat\theta-\theta )^2 -\frac{m}{2}(\hat\beta-\theta-\delta)^2 - \frac{\vert\delta\vert}{\hat b} \right\},$$
where $\hat b =\vert\hat\delta\vert (n+m)^{-\tau}$ is a data-adaptive choice of the scale parameter of a zero-mean Laplace prior, $\pi(\delta) = \left(2\hat b\right)^{-1}\exp\left(-\vert\delta\vert\hat b^{-1}\right)$. 
In Section \ref{sec:simulations}, Monte-Carlo simulations use $\tau = 0.25$, which determines the tuning parameter $(n+m)^{\tau}$.
\subsection{Bayes estimators with power prior [HDPP and EBPP]} 
\label{sec:bayes_estimators}
Popular Bayesian DIB methods include the power prior approach (\cite{ibrahim2000,ibrahim2015}), the (robust) meta-analytic approach (\cite{neuenschwander2010,schmidli2014}), and the commensurate prior approach (\cite{hobbs2012}). Power and commensurate priors discount external information by introducing a specific value or a distribution of the conflict parameter into the joint external and current data likelihood. For a normal outcome with a normal prior, analytic connections between the parameters of each approach have been shown for fixed, and, in certain cases, estimated values (see \cite{neuenschwander2020, wiesenfarth2019, pawel2022}). The robust mixture prior achieves DIB differently, i.e. by direct modification of the external analysis posterior/current analysis prior.  
The power prior is $\pi(\theta\vert\hat\beta,\gamma) = \pi_0(\theta) L(\theta;\hat\beta)^\gamma /C(\gamma)$, where $\gamma \in [0,1]$ determines the fraction of external data retained in the analysis and  $C(\gamma)=\int \pi_0(\theta) L(\theta;\hat\beta)^\gamma d\theta$ is a normalising constant. 

In our normal settings $L(\theta; \hat\beta)=\left[\sqrt{m}\phi(\sqrt{m}(\hat\beta - \theta))\right]$ and $L(\theta; \hat\theta)=\left[\sqrt{n}\phi(\sqrt{n}(\hat\theta - \theta))\right]$, where $\phi$ denotes the standard normal probability density function. The prior $\pi_0(\theta)$ is often chosen to be a normal distribution with very large variance, so that its impact on posterior inferences is minimized. After some minor algebra, the posterior density for a given $\gamma$ is
\begin{align*}
\pi(\theta \mid \hat\theta,\hat\beta,\gamma) &\propto \left[\sqrt{m}\phi(\sqrt{m}(\hat\beta - \theta))\right]^{\gamma}\cdot \left[\sqrt{n}\phi(\sqrt{n}(\hat\theta - \theta))\right]\\ &\propto \sqrt{n+m\gamma}\phi\left(\sqrt{n+m\gamma}\left(\frac{n\hat\theta + m\gamma \hat\beta}{n+m\gamma} -  \theta\right)\right).
\end{align*}
The posterior mean 
$=\frac{n\hat\theta + m\gamma \hat\beta}{n+m\gamma} = \hat\theta + \frac{m}{m + n/\gamma}\hat\delta$ and is equivalent to pooling $\hat\theta$ with variance $1/n$ and $\hat\beta$ with inflated by a factor $1/\gamma$ variance $1/(m\gamma)$. This posterior mean is equivalent to OMMSE when $\delta^2=(1-\gamma)/(m\gamma)$. 

When  $\delta=0$, the optimal $\gamma=1$, i.e., external information is fully borrowed and the posterior mean is equivalent to $\hat\theta^p$. Smaller $\gamma$ lead to fractional information borrowing. 

There are many ways to estimate $\gamma$. A similarity measure was suggested to estimate $\gamma$ in \cite{thompson2021dynamic}. This new measure is equal to the posterior probability that $\vert \delta \vert >0$ if $\hat\delta$ is below a clinically relevant threshold and $0$ otherwise. In
\cite{nikolakopoulos2018dynamic}, authors considered two more versions of the power parameter, which also use test-then-pool. 

Hellinger distance between normalized likelihoods associated with $\hat\theta$ and $\hat\beta$ can be used to estimate $\gamma$ (\cite{ollier2020}): 
$$H^2(\hat\beta,\hat\theta) =
\frac{1}{2} \displaystyle \int 
\left(\sqrt{\frac{L(\theta\vert\hat\theta)^{M_1}}{\int L(\theta\vert\hat\theta)^{M_1}d\theta}} - \sqrt{\frac{L(\theta\vert\hat\beta)^{M_2}}{\int L(\theta\vert\hat\beta)^{M_2} d \theta}}\right)^2 d \theta,$$ 
where $M_1=\min\left(1,\frac{m}{n}\right)$ and $M_2=\min\left(1,\frac{n}{m}\right)$. So that $\hat{\gamma}^{hd,pp}= \left(1-H(\hat\beta,\hat\theta) \right)^2$, which in our case is $\hat{\gamma}^{hd,pp} = \left(1-\sqrt{1-\exp\left(-n\hat\delta^2/8\right)}\right)^2$.
Then, a posterior mean estimate for Hellinger-distance based power prior (HDPP) is 
\begin{equation*}\label{hd}
\hat\theta^{hd,pp} = \hat\theta + \frac{m}{m + n/\hat\gamma^{hd,pp}}\hat\delta.
\end{equation*} 

\cite{gravestock2017adaptive} suggested empirical Bayes approach (EBPP) to estimate $\gamma$ by maximization of $L(\gamma;\hat\theta, \hat\beta)= \int \pi(\theta\vert\hat\beta,\gamma) L(\theta;\hat\theta) d\theta$, so that in our normal settings, $$\hat{\gamma}^{eb,pp}=\arg \max_{\gamma} L(\gamma;\hat\theta, \hat\beta) = \frac{1/m}{\max\{\hat\delta^2,1/n + 1/m\}-1/n}$$ and
\begin{equation}\label{eb}
\hat\theta^{eb,pp} = \hat\theta + \frac{m}{m + n/\hat\gamma^{eb,pp}} \hat\delta.
\end{equation} 

\subsection{Bayes estimators with a prior on the conflict [NP and LSTP]}\label{sec:bayesestimators}
A full Bayesian approach assigns prior on both, $\theta$ and $\delta$. We continue using a non-informative flat prior on $\theta$ but impose Gaussian and location-scale $t$ priors on $\delta$, respectively.

Following \cite{pocock1975}, the assumed normal prior (NP) centered at zero $\pi(\delta)=\sqrt{n}\phi(\sqrt{n}(0- \delta))$ leads to posterior 
\begin{align*}
\pi(\theta, \delta\vert \hat\theta,\hat\beta) &\propto \sqrt{n}\phi(\sqrt{n}(0- \delta)) \sqrt{m}\phi(\sqrt{m}(\hat\beta - \theta -\delta)) \sqrt{n}\phi(\sqrt{n}(\hat\theta - \theta)).
\end{align*}
From $\pi(\theta, \delta\vert \hat\theta,\hat\beta) \propto 
\pi(\theta \vert  \hat\theta,\hat\beta) \pi(\delta \vert \hat\theta,\hat\beta)$
the posterior mode is
\begin{align*}
\left(\hat\delta^{np},\hat\theta^{np}\right) &= \left( \frac{\hat\delta/(1/n+1/m)}{1/(1/n + 1/m) + n},\frac{\hat\theta n + \hat\delta/(1/n + 1/m)}{1/(1/n + 1/m) + n}\right)
 = \left( \frac{m \hat\delta}{2m+n}, \hat\theta + \hat\delta \frac{m}{2m+n}\right).
\end{align*}
Thus, the influence of external information on $\hat\theta^{np}$ is not reduced for large conflict.

A $t$ distribution prior fully discards prior information for large $\delta$ and Gaussian data (\cite{dawid1973,ohagan1979}). 
Assuming $\delta \sim lst(v,0,1/\sqrt{n})$, i.e., a location-scale $t$ prior centered at zero, with scale $1/\sqrt{n}$ and $v$ degrees of freedom,
the posterior becomes
\begin{align*}
\pi(\theta, \delta \vert \hat\theta,\hat\beta) &\propto \left[v+\left(\sqrt{n}\delta\right)^2\right]^{-(v+1)/2} \sqrt{m}\phi(\sqrt{m}(\hat\beta - \theta - \delta))  \sqrt{n}\phi(\sqrt{n}(\hat\theta - \theta)).
\end{align*}
This posterior requires numeric approximation to find posterior mode or mean. In our simulations, we use a location-scale $t$ prior with $v=3$, calculate the posterior mode and denote the estimator as LSTP.   

\subsection{Limited translation rules [LRT] estimators}\label{sec:LTR}

LTR estimators were suggested in \cite{efron1972limiting} as a compromise between a Bayes estimator and a frequentist minimax risk estimator. LTR estimators, akin to TTPool, rely on hypothesis testing. 
Adopting formulas (3.1)-(3.9) from \cite{efron1972limiting}, we use 
$\hat h = \sqrt{n} \hat \delta \sim N(\sqrt{n}\delta,1+n/m),$ and 
$h = \sqrt{n}\delta \sim N(0,1)$. Then, the posterior mean for the local conflict is $h^*=\hat{h}\frac{m}{2m+n}$ and the posterior mean for $\theta$ is equal to $\hat\theta^{np}$.
Thus, at $M=\sqrt{1/n+1/m}$ and $C=M(2m+n)/(m+n)$, the LTR estimator of $(\delta,\theta)$ is 
\begin{align*}
\hat\delta^{ltr} &= I\left(\hat\delta > C \right) \left(\hat\delta - M\right) +
I\left(\hat\delta < -C \right) \left(\hat\delta + M\right) + I\left(\vert \hat\delta\vert  \le C\right)\hat\delta^{np} \\
\hat\theta^{ltr} &= I\left(\hat\delta > C \right) \left(\hat\theta - M\frac{m}{m+n} \right) +
I\left(\hat\delta < -C \right) \left(\hat\theta + M \frac{m}{m+n}\right) + I\left(\vert \hat\delta\vert  \le C\right)\hat\theta^{np}.
\end{align*}

\subsection{Generalized DIB estimators [GDIB]} \label{sec:expanded} Among the considered estimators, TTPool, AMMSE, HDPP, EBPP, LSTP, and ALASSO suppress external information if $\delta$ is a fixed value different from zero and benefit from external information when $\delta=0$. 

ALASSO is the only estimator with the asymptotic oracle property. At the same time, as shown later in Theorem \ref{th6}, ALASSO's asymptotic power, by analogy with the pooled estimator, is equal to type 1 error against any local alternative: for any local alternative ALASSO estimator is asymptotically equivalent to the pooled estimator. 

For given $n$ and $m$, shapes of distributions of TTPool, AMMSE, HDPP and EBPP are fully determined by the distribution of $\sqrt{n}\hat\delta$, whereas the distribution of $\hat\theta$ only changes their locations; changes in $\theta$ shift the whole distribution. Using a sensitivity parameter, $s$, previously suggested in \cite{Tarima2013} to suppress or exuberate the effect of $\sqrt{n}\hat\delta$, the $\text{DIBs}$  from the set $\{\text{TTPool}, \text{AMMSE}, \text{HDPP and EBPP}\}$ are generalized to: \begin{align*}
\hat\theta^{GDIB}\left(g, s \right) &= \left[1-g\left(n\hat\delta^2s\right)\right]\hat\theta  + g\left(n\hat\delta^2s\right) \hat \beta = \hat\theta + g\left(n\hat\delta^2s\right) \hat \delta,
\end{align*}
where the function $g\left(n\hat\delta^2s\right)$ is an estimator of the mixing parameter $\lambda$ and defined on the unit interval. When the \textit{sensitivity-to-conflict} is tied to $n$ via   $s=n^{-2\tau}$at $\tau \in (0,0.5)$ it becomes asymptotically equivalent to the pooled estimator (see Theorem \ref{th4}). The function $g\left(n\hat\delta^2s\right)$ 
 is an estimator of  $\lambda \in [0,1]$ when $s\in [0,+\infty)$.
In $\hat\theta^{GDIB}(g,s)$, with $s=n^{-2\tau}$ a tuning parameter $\tau$ is introduced and $s$ depends on $n$. Both parameterizations, via $s$ ($n$- independent) and via $\tau$ ($n$-dependent)  describe departure from a base estimator, but asymptotic properties differ. For example, if $g\left(sn\hat\delta^2\right) = m/\left(n + m + m\cdot n\hat\delta^2\right) s$, the GDIB family is built around AMMSE (\cite{Tarima2013})
\begin{align*}
\hat\theta^{ammse}(s) = \hat\theta + \frac{m}{n + m + m\cdot n\hat\delta^2 s} \hat\delta.
\end{align*}
The $\hat\theta^{ammse}(s)$ used in our illustrative example for choosing $s$ to secure desired asymptotic properties.

\section{Large Sample Properties}
\label{sec:asymptotics} 

Asymptotics is considered at $n\to\infty$ with $\delta = h/\sqrt{n}$. We assume $n(n+m)^{-1} \to p$, $0 \le p\le 1$, [$1+n/m \to 1/(1-p)$ and $n/m \to p/(1-p)$], $\zeta_1 := \sqrt{n}\left(\hat\theta-\theta\right) \overset{d}{=} N(0,1)$, and $\zeta_2 
 := \sqrt{m}\left(\hat\beta-\beta\right) \overset{d}{=} N(0,1)$; $\zeta_1$ and $\zeta_2$ are mutually independent; $\xi := \sqrt{1-p}(h-\zeta_1) + \sqrt{p}\zeta_2$. 
 
\begin{theorem}\label{th1}
 $\sqrt{n}\left(\hat\beta-\theta\right) \overset{d}{\to} N\left(h,p/(1-p)\right)$.
\end{theorem}


\begin{theorem}\label{th2}
    $\sqrt{n}\left(\hat\theta^p-\theta\right) \to N\left((1-p)h,p\right)$.
\end{theorem}

\begin{theorem}\label{th3}
$$\sqrt{n}\left(\hat\theta^{ttp}-\theta\right) \overset{d}{\to} \text{Pr}(\xi^2>c) \zeta_1 + \text{Pr}(\xi^2 \le c) \left(p\zeta_1+\sqrt{p(1-p)}\zeta_2 + (1-p)h\right)$$
 \end{theorem}

Thus, the distribution of TTPool is an informative mixture of two dependent Gaussian random variables: $\zeta_1$ and $\zeta_p$ (see Appendix B). This mixture results in a unimodal distribution at $\delta=0$, then as $\delta$ increases bi-modality becomes clearly seen. As $\delta$ becomes even larger the contribution from the external dataset is suppressed and the density converges to the density of $\hat\delta$. When $\delta = 0$ and $c=3.84$, $\sqrt{n}\left(\hat\theta^{ttp}-0\right) \to 0.05 \cdot N(0,1) + 0.95 \cdot N\left(0,p^{-1}\right)$. If $\delta$ is a fixed $\ne 0$ value, $\text{Pr}\left(\hat\xi^2>c\right) \to 1$ and $\sqrt{n}\left(\hat\theta^{ttp} - \theta\right) \overset{d}{\to} N\left(0,1\right)$.

 \begin{theorem}\label{th4} If $\tau \in (0,0.5)$, $\sqrt{n}\left(\hat\theta^{alasso}_n(\tau)-\theta\right)$ and $\sqrt{n}\left(\hat\theta^{GDIB}_n\left(g,n^{-2\tau}\right)-\theta\right)$ converge to $N\left((1-p)h,p\right)$ in distribution.
 \end{theorem}
\begin{theorem}\label{th5} $\sqrt{n}\left(\hat\theta^{GDIB}_n\left(g, s\right)-\theta\right) \overset{d}{\to} \zeta_1 + g\left(\frac{s\xi^2}{1-p}\right) \xi.$
 \end{theorem}
 \begin{corollary}\label{cor1}   $\sqrt{n}\left(\hat\theta^{eb}-\theta\right) \overset{d}{\to} \zeta_1 + \frac{1-p}{1-p+p/\gamma^{eb}}\xi$, $\sqrt{n}\left(\hat\theta^{hd}-\theta\right) \overset{d}{\to} \zeta_1 + \frac{1-p}{1-p+p/\gamma^{hd}}\xi$,  
 where $\gamma^{eb}:= p\left(\max\{\xi^2,1\}-1+p\right)^{-1}$ and $\gamma^{hd} := \left(1-\sqrt{1-\exp\left(-\xi^2/(8-8p)\right)}\right)^2$.
 \end{corollary}

AMMSE asymptotics is found in \cite{van2023rejoinder} and follows from Theorem \ref{th5}:
 \begin{corollary}[AMMSE\label{cor2}] $\sqrt{n}\left(\hat\theta^{ammse}-\theta\right) \overset{d}{\to} \zeta_1 + \frac{\sqrt{1-p}}{1+\xi^2}\xi$\end{corollary}

No asymptotics was derived for NP, LSTP and LRT: NP and LRT do not suppress external data if $\delta \ne 0$ (Section \ref{sec:simulations}). Derivation of LSTP distribution is analytically challenging. Numeric approximation was used to find LSTP's posterior mode. Section \ref{sec:simulations} shows that conflicting external data is eventually suppressed  by LSTP. Heavy tailed priors are especially important for practical applications of fully Bayes DIB.

\section{Simulations}
\label{sec:simulations} 
Section \ref{sec:mse} uses $\theta = 0$; Section uses \ref{sec:hypothesis.testing} $\theta \in \{0.03, 0.06, 0.09\}$. The current sample is summarized by $\hat\theta \sim N(0,1/n)$, $n=1,000$. The external data is given by $\hat\beta \sim N(\theta +\delta,1/m)$, $m=100,000$. Appendix B shows DIB densities at $\sqrt{n}\delta \in \{0,0.32, 1.58, 5.06\}$. 

\subsection{Mean Squared Error} \label{sec:mse}

\begin{figure*}[t]
\centering
\includegraphics[width=5.8in]{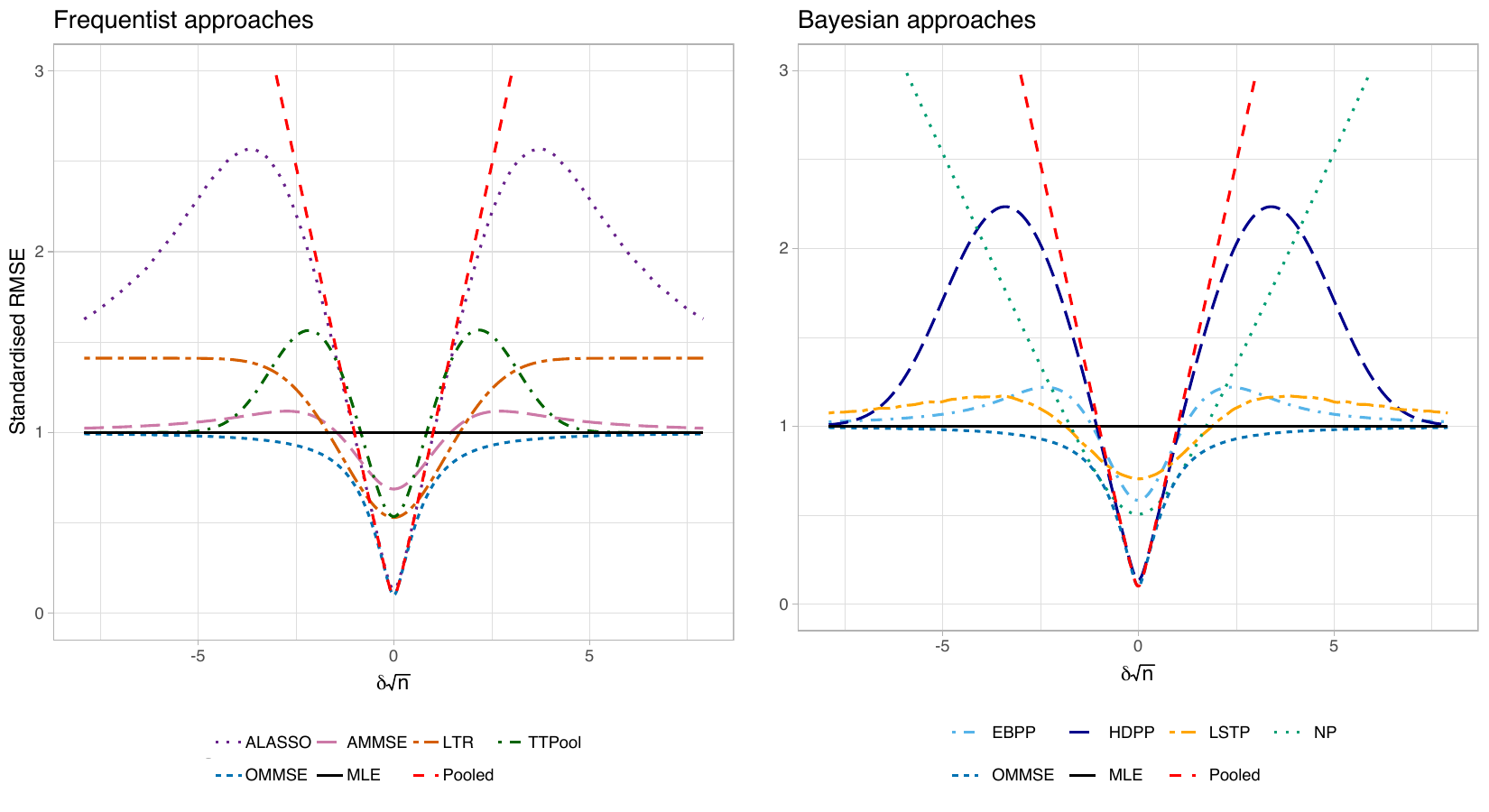}
\caption{Standardized root mean squared error for Bayesian and frequentist borrowing approaches for varying conflict $\delta \sqrt{n}$. Results are derived from grid-based numerical integration.}
\label{fig:MSEs}
\end{figure*}

Figure \ref{fig:MSEs} reports SRMSE (Standardized Square Root of MSEs), which we define as $\sqrt{n\cdot MSE}$. Since MSE of the MLE is $1/n$, its SRMSE is equal to one. Then, SRMSE can be viewed as a square root of the relative efficiency defined as the ratio of MSE of a DIB estimator and MSE of the MLE. Patterns of SRMSEs in Figure \ref{fig:MSEs} are similar to what was reported by others (e.g., \cite{yu2015adaptive,nikolakopoulos2018dynamic,calderazzo2023coping}), but we explore them jointly with respect to the lower bound determined by OMMSE. SRMSE approximates the limiting risk of $\theta$ estimation for a quadratic loss.


To determine an optimal DIB estimator a single optimality criterion is needed. One option would be to assign a cost for each value of $\delta$. From the Bayes perspective this cost is naturally determined by a prior, $\pi(\delta)$. Then, the integrated over $\pi(\delta)$ MSE is
\begin{align*}
IMSE(\hat\theta^{DIB}\vert \theta, \pi) &= \int_{-\infty}^{+\infty} MSE(\hat\theta^{DIB}\vert \theta, \delta) \pi(\delta)d\delta \\&= \int_{-\infty}^{+\infty} \int_{-\infty}^{+\infty} (z - \theta)^2   f_{\hat\theta^{DIB}}(z\vert \theta, \delta) \pi(\delta) dz d\delta,
\end{align*}
where $f_{\hat\theta^{DIB}}(z\vert \theta, \delta)$ refers to the sampling distribution of a DIB estimator.

TTPool, NP, AMMSE, HDPP, EBPP, and Generalized DIB are \textit{location invariant} with respect to $\theta$: any change in $\theta$ is associated with a shift of a whole distribution of a DIB estimator. The location invariance property of DIB estimators ensures that the integrated MSE is also invariant to changes in $\theta$: $IMSE(\hat\theta^{DIB}\vert \theta, \pi) = IMSE(\hat\theta^{DIB}\vert \pi)$. 

\begin{table}[ht]
    \centering
    \begin{tabular}{c|ccccc}
Estimator & $\pi_1(\delta)$ & $\pi_2(\delta)$ & $\pi_3(\delta)$ & $\pi_4(\delta)$ & $\pi_5(\delta)$\\ 
\hline
  MLE & 1.00 & 1.00 & \textbf{1.00} & 1.00 & 1.00 \\ 
  Pooled & 0.80 & 1.38 & 15.66 & 0.58 & 1.10 \\ 
  NP & \textbf{0.68} & 0.91 & 7.91 & 0.61 & 0.81 \\ 
  AMMSE & 0.83 & 0.92 & 1.01 & 0.78 & 0.86 \\ 
  TTPool & 0.94 & 1.12 & 1.03 & 0.80 & 0.98 \\ 
  ALASSO & 0.78 & 1.23 & 1.35 & 0.57 & 0.94 \\ 
  EBPP & 0.82 & 0.95 & 1.02 & 0.74 & 0.86 \\ 
  HDPP & 0.75 & 1.14 & 1.12 & \textbf{0.55} & 0.87 \\ 
  LTR & 0.71 & \textbf{0.87} & 1.36 & 0.64 & \textbf{0.77} \\ 
  LSTP & 0.80 & 0.89 & 1.02 & 0.76 & 0.83 \\ 
  \hline
OMMSE & 0.53 & 0.67 & 0.97 & 0.42 & 0.58  \\ 
\end{tabular}
\caption{Bayes risk at different choices of $\pi(\delta)$: $\pi_1(\delta)=N(0,1/n)$, $\pi_2(\delta)=N(0,3/n+3/m)$, $\pi_3(\delta)=Unif(-1,1)$, $\pi_4(\delta) = Laplace(0,(n+m)^{-0.35})$, $\pi_5(\delta) = LST(3,0,1/\sqrt{n})$; LST stands for a location scale T distribution. Results are derived from grid-based numerical integration.}
\label{tab:Risk}
\end{table}

\begin{figure*}[t]
  \begin{center}
 \includegraphics[width=1\linewidth]{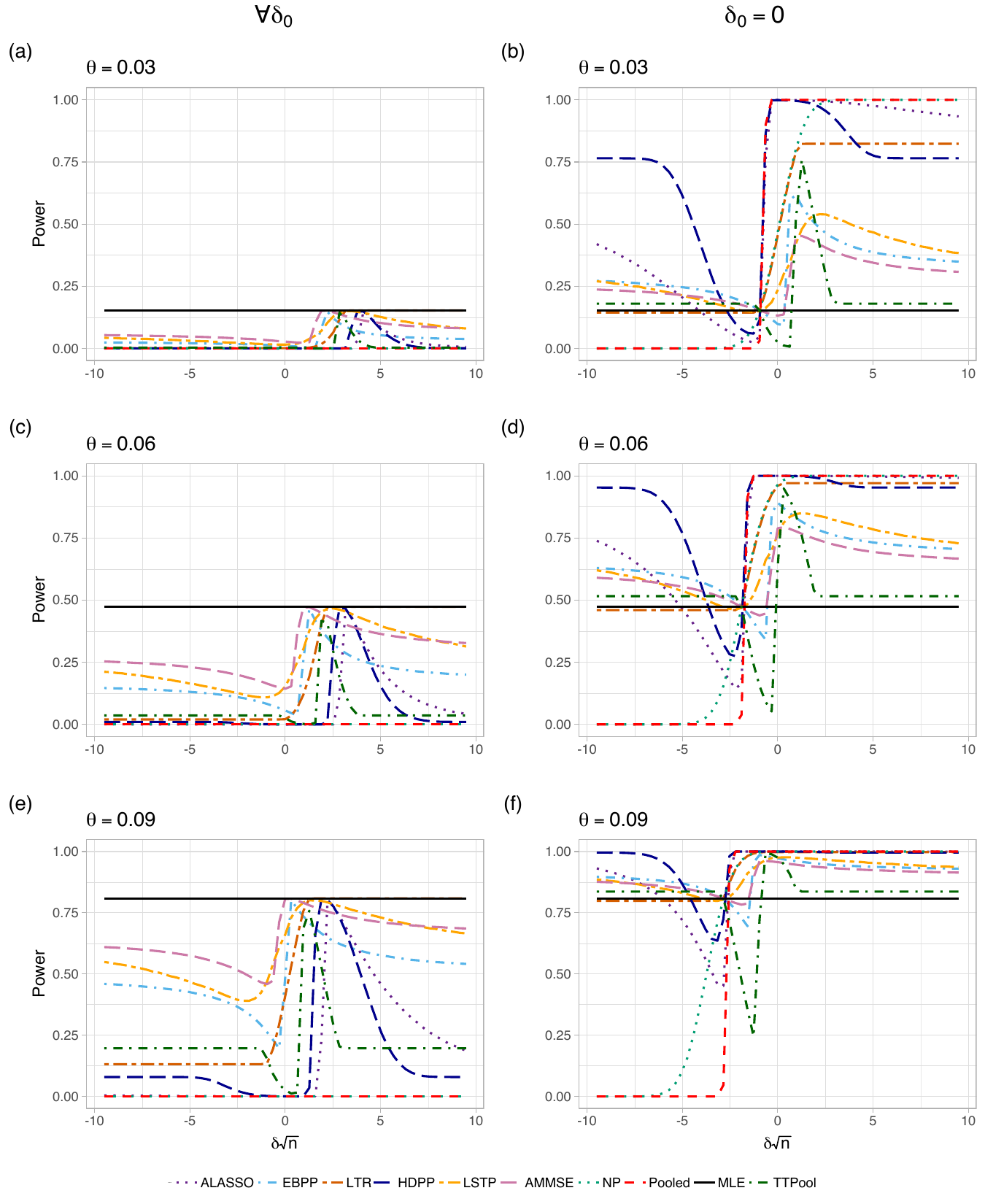}
  \caption{\label{fig:crit.vals} Power as a function of $\delta$ at different $\theta$ and different definitions of $H_0$; T1ER $= 0.025$. Results are derived from grid-based numerical integration.}
  \end{center}
\end{figure*}

Note that $$IMSE(\hat\theta^{DIB}, \pi) = E_{\hat\theta^{DIB}} \left(E_{\pi} (z-\theta)^2\vert \hat\theta^{DIB} = z\right) \ge E_{\hat\theta^{*}} \left(E_{\pi} (z-\theta)^2\vert \hat\theta^{*} = z\right),$$ where $\hat\theta^{*}$ is the Bayes optimal decision (see Section 8.8 \cite{degroot2005optimal}). In Section 8.9 \cite{degroot2005optimal}, the optimal decision is obtained by minimizing an expected posterior loss for a given $\mathbf{X}_n$ and $\mathbf{Y}_n$
In the case of quadratic loss, the expected posterior is minimized by the posterior mean. 
Thus, the IMSE is minimized when the DIB is the posterior mean under the prior $\pi(\delta)$ (see e.g. \cite{parm2009}).

Table \ref{tab:Risk} reports a few versions of IMSE calculated for the DIB estimators. For computational simplicity, our simulations use the posterior mode as LSTP estimator, which does not guarantee optimality in terms of IMSE minimization because it may not coincide with the posterior mean.
\begin{figure*}[t]
  \begin{center}
 \includegraphics[width=1\linewidth]{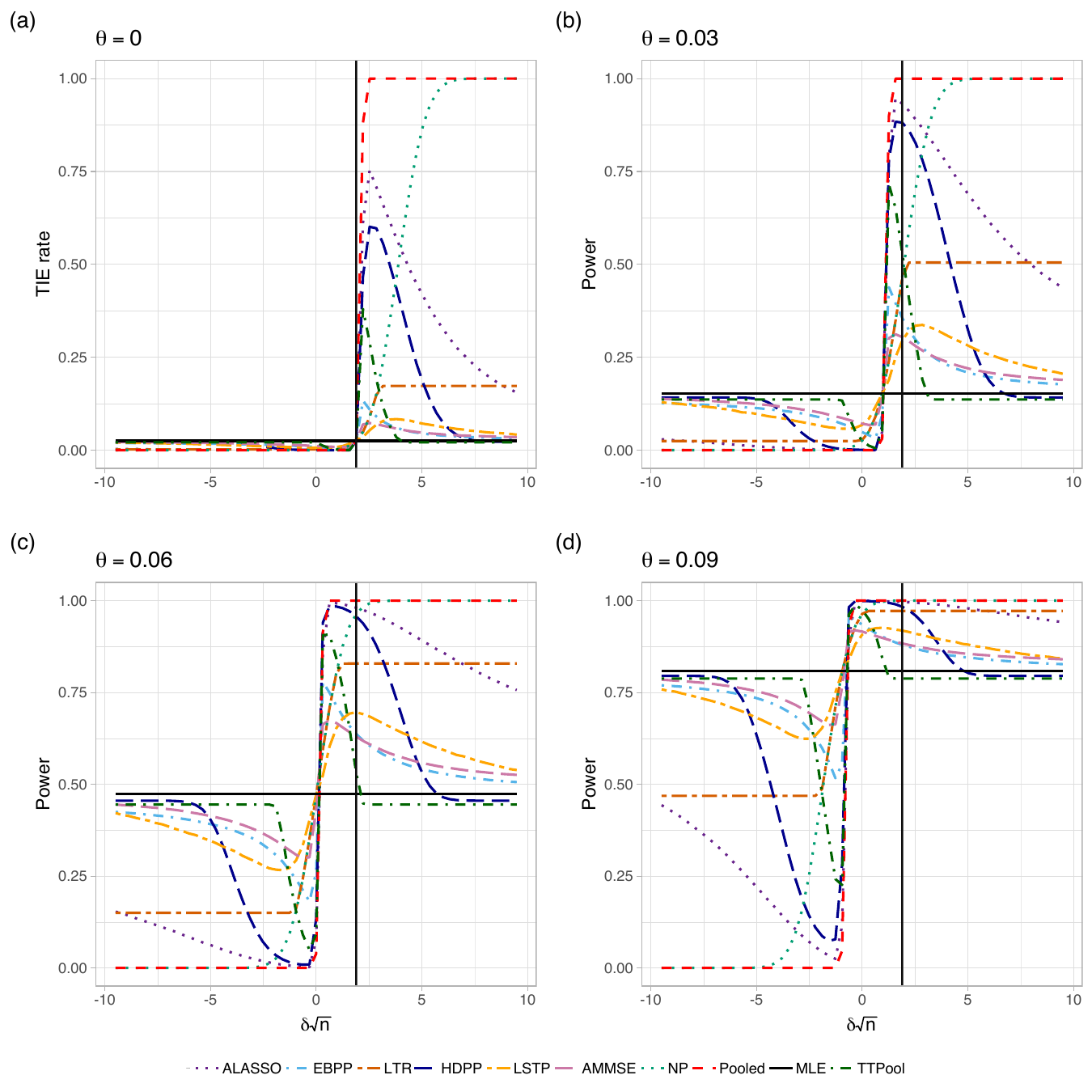}
  \caption{\label{fig:power2} Type I error rate and power as a function of $\delta$ at different $\theta$, with critical value obtained assuming $\delta < 0.06$ under $H_0$. The vertical line is at $\delta\sqrt{n}=0.06\sqrt{1000}$. Results are based on $5\times 10^4$ Monte Carlo samples.}
  \end{center}
\end{figure*}
\subsection{Hypothesis testing} \label{sec:hypothesis.testing}
As shown in Section \ref{sec:estimators} and illustrated in Web Appendix B, majority of the DIB estimators which can be used for hypothesis testing have different from normal distributions in the local neighbourhood of $\delta=0$. Consequently, choices of their critical values depend on $\delta$. Figures \ref{fig:crit.vals}(a-b) [row one], \ref{fig:crit.vals}(c-d) [row two], and \ref{fig:crit.vals}(e-f) [row three] show power properties when $\theta\in\{0.03, 0.06, 0.09\}$, respectively. For each scenario T1ER is secured at 2.5\% type, but the null hypotheses is defined differently for each column of Figure \ref{fig:crit.vals}. 

Column one represents the most conservative situation when $\forall\delta$, T1ER is controlled exactly. The problem with such setting is that statistical power is the highest if the testing is based on $\hat\theta$ (the black horizontal line). These simulations confirm the findings of \cite{kopp2020power} and \cite{kopp2024}, where the authors showed that no power gain is possible when external information is incorporated in a statistical procedure if a uniformly most powerful test exists. Column one also indicate that there exist just a single value of $\delta$ where AMMSE, ALASSO, LTR, EBPP, LSTP, and HDPP estimators show comparable to the MLE power. By the Karlin-Rubin lemma the likelihood ratio test (equivalent to the MLE in our settings) is uniformly ($\forall H_A:\theta=\theta_A$) most powerful if data came from the exponential family.

With increasing sample size ALASSO's the power to detect local departures from $H_0:\theta = 0$ is equal to zero, because ALASSO procedure forces all local ALASSO estimates to zero.
\begin{theorem} \label{th6}
Asymptotic power to detect $H_0:\theta = h_{\theta}/\sqrt{n} $ is equal to zero if testing is based on $\hat\theta_n^{p}$, $\hat\theta_n^{alasso}(\tau)$ or $\hat\theta_n^{DIB}\left(g,n^{-2\tau}\right)$ at $\tau\in (0,0.5)$.
\end{theorem}

T1ER control $\forall \delta$ is an extreme setting: information borrowing becomes useless.

Another extreme is to determine the critical value for testing $H_0:\theta = 0$ assuming $\delta = 0$, see the last column of Figure \ref{fig:crit.vals}. Statistical powers when $\theta\in\{0.03, 0.06, 0.09\}$ are shown in Figures \ref{fig:crit.vals}(b,d,f), respectively. In this setting, the DIB estimators are the most useful for testing $\theta=0$ and the pooled estimator tops them all in terms of statistical power. 

Finally, researchers can consider a compromise when the critical value is chosen to be the highest critical value within a range of plausible departures of the conflict from zero. Figure \ref{fig:power2} reports T1ER and power when the critical value is selected to control T1ER for testing $H_0: \theta=0$ for $\delta < \delta_0 = 0.0636$. The threshold $\delta_0$ is chosen by an investigator independently of the data and heavily depends on a subject area.

Figure \ref{fig:power2}(a) shows that T1ER is exactly controlled when $\delta = \delta_0$ across all methods, and can be only smaller when $\delta < \delta_0$. The pooled estimator and NP based testing do not suppress incorrect external information  can have T1ER as high as 100\%. T1ER of the test based on the LTR estimator converges to a constant value as a conflict increases. 

Small positive departures from $\theta=0$ (Figure \ref{fig:power2}(b)) show similar power properties for all DIBs when $\theta = \delta_0/2$ and $\delta = \delta_0/2$. The range $\delta_0/2 < \delta \le \delta_0$ at $\theta = \delta_0/2$ shows higher power properties when a DIB is used for testing, but $\delta_0/2 < \delta \le \delta_0$ at $\theta = \delta_0/2$ reduces power as compared to the MLE. 

Overall other power curves (Figure \ref{fig:power2}(c,d)) show similar patterns to Figure \ref{fig:power2}(b). When $\delta < \delta_0$, T1ER for testing $\theta$ is controlled at a predetermined level, and as simulations show there is a sub-range of values of $\delta$ ($\delta \in R = (\delta_0 - \theta, \delta_0)$) where the power of detecting $\theta>0$ is higher than the power of test based on the MLE. This is analogous to the ``sweet spot'' region of \cite{viele2014use}, which is determined by MSE reduction as compared to the MSE of the MLE. 

Due to the complexity of distributions of the DIB estimators and non-monotonicity of their power curves, the region $R$ where T1ER is controlled and the power of superiority testing is higher than the power of the MLE may need to be determined numerically.

Thus, if a researcher is willing to benefit from using a DIB estimator for hypothesis testing, some additional specifications on the conflict should be in place. 

Control of the additional specifications on the conflict, e.g. $\delta \le \delta_0$, in real-life settings may be problematic, and researchers may be tempted to probabilistically describe how likely these additional specifications hold true. For example, a researcher can opt for reporting an estimate of $p_2=Pr\left(\sqrt{n}\delta_0 - \sqrt{n}\hat\theta < \sqrt{n}\hat\delta < \sqrt{n}\delta_0\right)$ to describe how likely the conflict is in the ``sweet spot'' (T1ER is controlled and power is higher than the power of the MLE) and/or $p_3=Pr\left(\sqrt{n}\hat\delta < \sqrt{n}\delta_0\right)$ to describe how likely $\delta$ belongs to the range where T1ER is controlled. However, $p_2 \le p_3 \le \text{Pr}\left(\sqrt{n}\hat\theta < \sqrt{n}\delta_0\right)$ under $\theta=0$, where the last inequality  is a result of $var(\sqrt{n}\hat\delta) \ge var(\sqrt{n}\hat\theta)$. Then, under $\sqrt{n}\delta_0=1.96$, $\text{Pr}\left(\sqrt{n}\hat\theta < 1.96\vert\theta=0\right) = 0.975$ leads to $\text{Pr}\left(\sqrt{n}\hat\delta < 1.96\vert \theta=0\right) \le 0.975$. So, T1ER of the test $\{\sqrt{n}\hat\theta > 1.96\}$ for testing $\theta=0$ is always smaller than T1ER of the test $\{\sqrt{n}\hat\delta > 1.96\}$ for testing $\delta=0$. This controversy can be formulated as follows: the probability that T1ER is not controlled, $1-p_3=\text{Pr}\{\sqrt{n}\hat\delta > 1.96\}$, is always higher than T1ER of the MLE-based test for all common sense $\delta_0$ ($\sqrt{n}\delta_0$ above a critical value, e.g. $1.96$, are of no interest for testing $H_0:\theta=0$). Thus, we emphasize the importance of choosing $\delta_0$ using expert opinion and common sense. Sample estimates of $p_2$ and $p_3$ can be explored as functions of $\delta_0$ in a sensitivity analysis as is completed in Section \ref{Illustrative_Example1}, but their practical relevance is questionable. 

\textbf{Remark:} Theorem \ref{th6} uses a sample size dependent definition of the sensitivity-to-conflict parameter $(s=n^{-2\tau})$, which makes the sensitivity parameter converge to zero as sample size increases, which ensures convergence of $\hat\theta_n^{DIB}\left(g,n^{-2\tau}\right)$ to the pooled estimator. Theorem \ref{th6} does not apply to GDIB estimators with a sample size independent $s>0$.

Note that in this section $\theta_0=0$ is assumed; in Section \ref{Illustrative_Example1}, $\theta_0 \neq 0$ and $\theta$ needs to be substituted with $\theta-\theta_0$ to make the formulas of this section applicable.  
\section{Illustrative Example}\label{Illustrative_Example1}
The Pregnancy Risk Assessment Monitoring System (PRAMS) began in 1987 as a response to a continuing high rate of maternal and infant mortality in the United States. PRAMS is conducted by the Centers for Disease Control and Prevention’s Division of Reproductive Health in collaboration with state health departments 
PRAMS continues today as a state-based surveillance system (e.g., WIPRAMS for WI) of maternal behaviors, attitudes, and experiences before, during, and shortly after pregnancy (\cite{shulman_pregnancy_2018}). 

Using WIPRAMS, we estimated the mortality rate (=37/94=39.36\%) for extremely preterm infants (gestational age $<$28 weeks). 
We also know a 2017 USA-level finding the mortality 
is = 38.4\% ($m=20,000$). Sample SDs for WI and US rates are $SD_{wi} \approx 0.4886$ and $SD_{us} \approx 0.4864$, which lead to the standardized quantities: $\hat\theta_{st} = \hat\theta /SD_{wi}= 0.7895$ and $\hat\beta_{st}= \hat\beta /SD_{us} = 0.8057$. To borrow information 
we considered the family $\hat\theta^{ammse}(s)$.  

\begin{figure}
\centering
\begin{subfigure}{(a)}
\includegraphics[width=3.0in]{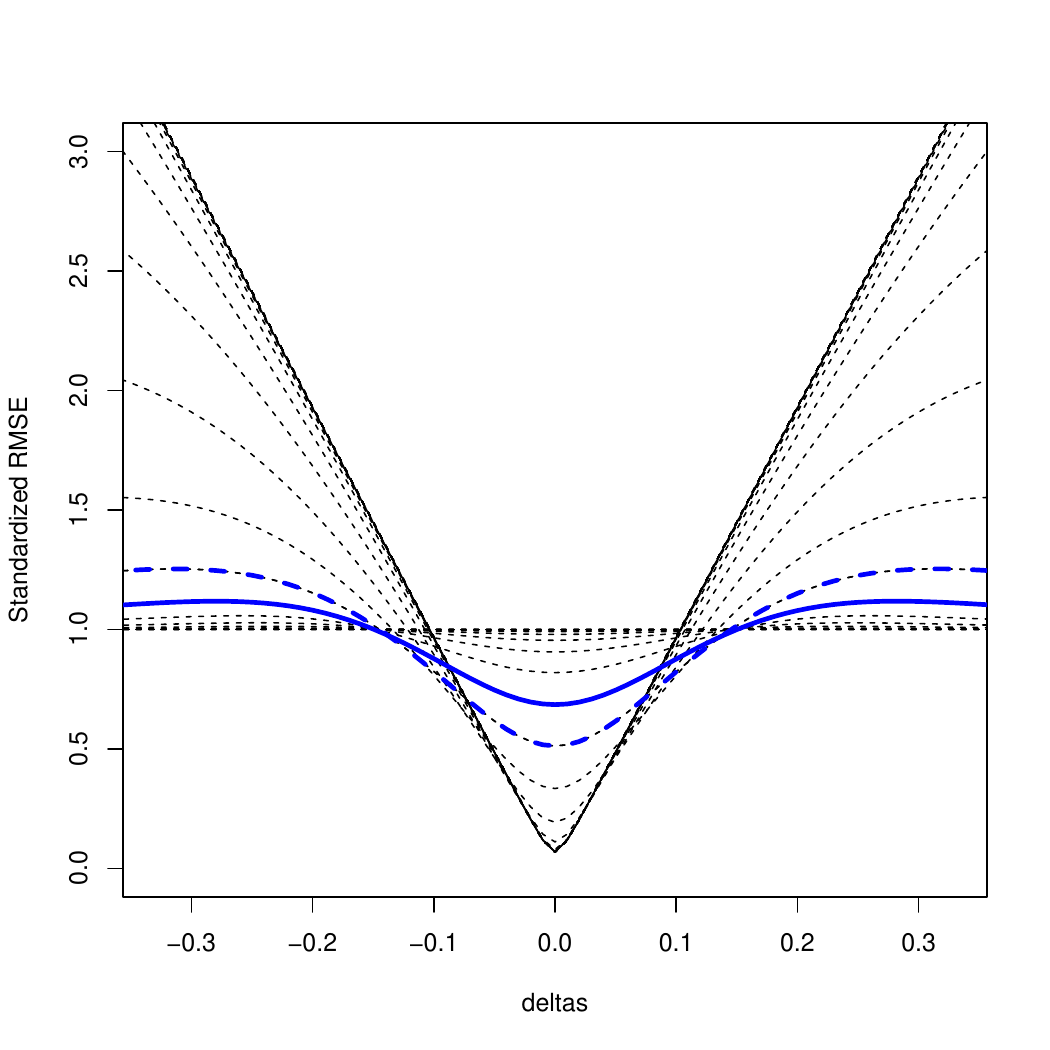} 
\end{subfigure}
\begin{subfigure}{(b)}
 \includegraphics[width=3.0in]{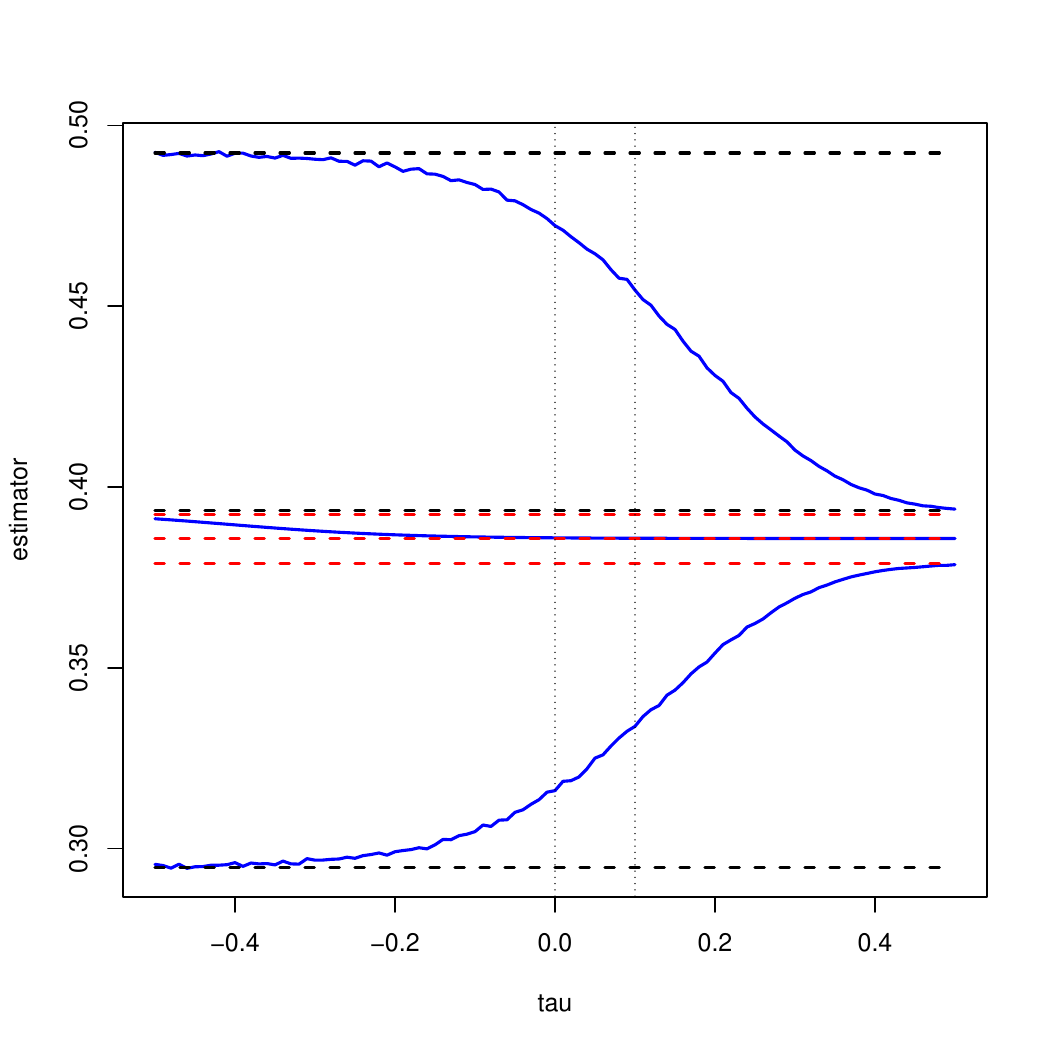}
\end{subfigure}
\caption{\label{fig_ex4}(a) SRMSE of $\hat\theta^{ammse}_n\left(n^{-2\tau}\right)$ for $\tau\in [-5,5]$; solid blue is for $\tau=0 (s=1)$; solid dashed blue is for $\tau=0.1 (s\approx 0.4)$. (b) Infant Mortality and 95\% confidence intervals; $\hat\theta$ (black dotted, $n=94$); $\hat\beta$ (red dotted, $m=20,000$); $\hat\theta^{ammse}_n\left(n^{-2\tau}\right)$ (blue)}
\end{figure}

After exploring dependence of $SRMSEs$ of $\hat\theta^{ammse}(s)$ $(s=n^{-2\tau})$ on $\delta$ for different values of $\tau$ (Figure \ref{fig_ex4}a), it was determined that $\tau = 0.1  (s=94^{-0.2} \approx 0.4)$ gives a reasonable compromise between maximum expected loss (highest MSE inflation is 25\% in comparison to the MSE of the MLE) and maximum expected benefits (about 50\% reduction of MSE if $\delta=0$). This choice of $s$ is not based on the observed data but uses Theorem \ref{th5} for asymptotic distribution of $\hat\theta^{ammse}(s)$. Then, using the observed data at $s = 0.4$, $\hat\theta^{ammse}(0.4) = 0.396$ with a 95\% bootstrap confidence interval ($10^7$ resamples) $0.334-0.454$. Figure \ref{fig_ex4}b shows confidence intervals for other choice of $\tau$.

For illustrative purposes, we also consider a hypothesis testing problem. Can we make a claim that \textit{more than two thirds of extremely preterm infants are alive at their first birthday}? This claim (our research hypothesis) needs further specifications imposed on $\delta$:

\textbf{Option 1 ($\forall \delta$):} As shown in Section \ref{sec:hypothesis.testing}, testing $H_0:\theta=1/3 (\forall \delta)$ versus $H_A:\theta<1/3 (\forall \delta)$ cannot benefit from information borrowing. Then, the MLE based testing relies on $Z_1=\sqrt{n}\left(\hat\theta-1/3\right) \sim N(0,1)$ under $H_0$. The p-value, $\text{Pr}\left(Z_1 > z_1\vert \theta=1/3\right) = 0.1157$, where the observed value of $Z_1$, $z_1=\sqrt{94}\left(0.8057-1/3\right)=1.1963$. Then, $H_0$ is not rejected.

\textbf{Option 2 ($\delta=0$):} We test $H_0: \theta=1/3$ under $\delta=0$: uncertainty about $\theta$ becomes  small and $H_0$ is rejected with $\hat\theta^{p}$. Using asymptotics of $Z_2 = \sqrt{n}\left(\hat\theta^{p}-1/3\right)$ under $H_0$ with $\delta=0$, $\text{Pr}\left(Z_2 > z_2 \vert \theta=1/3, \delta = 0\right) < 0.0001$, where $Z_2=z_2$. 

\textbf{Option 3  ($\delta\le \delta_0$):} $H_0: \theta=1/3(\delta  < \delta_0)$. 
Previously, we chose $\tau=0.1$, and to fully specify the hypothesis testing task, we choose $\delta_0=0.05$ and using Monte-Carlo simulations of $Z_3=\sqrt{n}\left(\hat\theta^{ammse}(0.4)-\theta\right)$ under $\theta=1/3$ and $\delta_0=0.05$. The p-value $\text{Pr}(\hat\theta^{ammse}(0.4)>z_3 \vert \delta=0.05, \theta=1/3)=0.0423$, where $z_3=1.0421$; $H_0$ is rejected, at $\delta \le \delta_0$.

The choice of $\delta_0=0.05$ was driven by common sense considerations and an expert opinion that it was highly unlikely that in the population of extremely preterm births, infant mortality rate in WI differs from the US by more than 5\%. If we consider $\delta_0=0.01$, p-value decreases to $0.0355$, whereas $\delta_0=0.087$ determines the tipping point when p-value$=0.05$ and statistical significance disappears for $\delta>0.087$. Sample estimates of $p_3=\text{Pr}(\hat\delta < \delta_0)$, $\hat p_3 = 0.60$ if $\delta_0 =0.01$, $\hat p_3 = 0.74$ if $\delta_0 =0.05$, and $\hat p_3 = 0.14$ if $\delta_0 =0.087$, are not overly useful because, as it was mentioned earlier, uncertainly associated with estimation of $\delta$ is always higher than uncertainty associated with estimation of $\theta$. Sample estimates of the probability of being in the sweet spot $p_2$ are even smaller than the estimates of $p_3$: $\hat p_2 = 0.60$ if $\delta_0 =0.01$, $\hat p_2 = 0.55$ if $\delta_0 =0.05$, and $\hat p_2 = 0$ if $\delta_0 =0.087$. Again, estimates of $p_2$ are of limited use. 


\section{Summary} \label{sec:summary}
This manuscript explored large sample properties of several DIB estimators: TTPool (test-then-pool), AMMSE (adaptive minimim mean squared error), ALASSO (adaptive lasso), HDPP (Hellinger distance based power prior), EBPP (empirical bayes power prior), NP (normal prior), LSTP (local-scale t prior), and LTR (limited translation rule). Their properties were evaluated in the local asymptotic framework where both the treatment effect $\theta$ and the conflict between the external and current datasets $\delta$ were evaluated in their local neighbourhoods: $\theta = h_{\theta}/\sqrt{n}$  and $\delta = h/\sqrt{n}$. Asymptotic distributions of TTPool, AMMSE, ALASSO, HDPP, and EBPP estimators are found to be non-Gaussian. As $h$ changes, DIB estimators are ``floating'' between two extremes: the pooled estimator, which fully borrows external information, and the MLE derived on the current data only. ALASSO, HDPP and EBPP can be viewed as empirical Bayes estimators, where prior parameters are estimated from the data. NP and LSTP are full Bayes estimators. AMMSE is fully frequentist estimator which an estimated version of the OMMSE - the estimator with the lowest possible MSE. TTPool is an estimator based on hypothesis testing and LRT is a compromise estimators between the full Bayes estimator pooling data from both data sources and the MLE. 

There are no DIB estimators with uniformly smallest MSE $\forall\delta$. However, if the MSE is integrated over some prior distribution $\pi(\delta)$ assigning prior weights to $\delta$, the optimization problem is fully determined and Bayes estimators obtained under $\pi(\delta)$  are optimal as they minimize the integrated MSE. There are some desired properties for $\pi(\delta)$: \textbf{(1)} $\pi(\delta)$ should suppress external data if $\delta \ne 0$, \textbf{(2)} a DIB should benefit from external information if $\delta=0$ or small, \textbf{(3)} limiting risk against local alternatives (the SRMSE in our setting) should be bounded. For example, pure Bayes NP estimator does not suppress external data so that its SRMSE diverges to infinity when the conflict increases. Property 1 also does not hold for the LRT estimator because its limiting MSE is always higher than the MSE of the MLE at large $\delta$. The LSTP (location scale $t$ distribution; location $=0$, scale $=1/n$, and degrees of freedom $=3$) preserves these three desirable properties: its heavy tails (3+ degrees of freedom ensure existence of the first two moments) suppress external data at high $\delta$ and the use of scale $=1/n$ incorporates external information at small $\delta$ at \textit{any} sample sizes. The main challenge associated with LSTP is computational intensity of calculating posterior means or modes. Variance of posterior mean with heavy tailed priors may be higher than variance of the MLE (\cite{peri1992}), which can be confusing to practitioners. 

Empirical Bayesian estimators (EBPP, HDPP and ALASSO) approximate optimal Bayes solutions. Using large sample properties one can evaluate dependence of SRMSE on $h$. ALASSO does not preserve Property 3:  $\forall h\ne0$ its SRMSE diverges as $n\to\infty$. Table \ref{tab:properties} summarizes the properties of DIB methods.

ALASSO is the only DIB estimator with the oracle property. ALASSO is asymptotically equivalent to the pooled estimator for finite $h$ and is asymptotically equivalent to the current data MLE when $\delta$ is a fixed sample size-independent value different from zero. When SRMSE is evaluated at a local conflict $\delta = h/\sqrt{n}$, $h\ne 0$, the estimator is asymptotically equivalent to the pooled estimator and its SRMSE is unbounded. Consequently, asymptotically, in contrast to many other DIB approaches (EBPP, HDPP, AMMSE, TTP, LST, GDIB$(g,s)$, and LSTP) ALASSO does not distinguish between $\delta = 0$ and $\delta = h/\sqrt{n}$ and opts for full information borrowing. 

Asymptotics of EBPP, HDPP, AMMSE and TTP is determined by $h_{\theta}$ and $h$. Sufficient statistic is two dimensional, $\sqrt{n} \hat\theta$ and $\sqrt{n} \hat\delta$, but DIB estimators are one-dimensional quantities. Then, optimal inference about $\theta$ is only possible when $\delta$ is known. This is why OMMSE sets MSE's lower bound, but cannot be applied in practice. It is still instructive to compare MSEs of DIB estimators against the lower bound.
 
 When $\sqrt{n} \hat\delta$ is replaced with $n^{-\tau} \sqrt{n} \hat\delta$, $\tau \in (0,0.5)$ (the sensitivity-to-conflict $s=n^{-2\tau}$ depends on $n$), the GDIB become asymptotically equivalent to ALASSO and enjoy the oracle property. By analogy with ALASSO at a fixed $h \ne 0$ conflicting external information is not asymptotically suppressed. To avoid the asymptotic impact of functional dependence of the sensitivity-to-conflict parameter ($s$) on sample size $\left(s=n^{-2\tau}\right)$, a sample size independent version of $s \ge 0$  is considered and $\sqrt{sn} \hat\delta$ is used instead of $\sqrt{n} \hat\delta$ in EBPP, HDPP, AMMSE and TTP. This modification changes local asymptotic properties without imposing the oracle property. Our illustrative example chose $s$ to determine an acceptable MSE profile independently of $n$, $\hat\theta$, and $\hat\delta$. 
\begin{table}[ht]
    \centering
    \begin{tabular}{c|c|c|c|c}
Estimator & Pr. 1 & Pr. 2 & Pr. 3 & Oracle Pr. \\ 
\hline
  MLE      &           &            & X          & \\ 
  Pooled   &           &  X         &            & \\ 
\hline
  NP       &           &  X         &            & \\ 
  AMMSE    & X         &  X         & X          & \\ 
  TTPool   & X         &  X         & X          & \\ 
  ALASSO   & X         &  X         &            & X \\ 
  EBPP     & X         &  X         & X          & \\ 
  HDPP     & X         &  X         & X          & \\ 
  LTR      & partially &  X         & X          & \\ 
  LSTP     & X         &  X         & X          & \\ 
  GDIB$\left(g,n^{-2\tau}\right)$ & X         &  X         &            & X \\
  GDIB$(g,s)$ & X         &  X         &     X       & \\
  \hline
OMMSE & X & X & X & X \\ 
\end{tabular}
  \caption{Property 1 (suppresses external data if conflict is present and large); Property 2 (benefits from external information is the conflict is absent or small); Property 3 (bounds limiting risk against local alternatives); Oracle Property (suppresses external data if conflict is present and large and fully uses external information is the conflict is absent}
    \label{tab:properties}
\end{table}

Hypothesis testing is problematic with DIB if one wishes to control T1ER $\forall\delta$, but under certain data-independent assumptions on $\delta$ DIB benefits can be substantial. In our illustrative example, superiority testing on $\theta$ under the assumption that $\delta < \delta_0$ increased statistical power while T1ER was still controlled. Within the local asymptotic framework, stochastic uncertainty that $\delta$ belongs to a desired region never goes away as $n\to\infty$ and data independent justification of the assumptions on $\delta$ are necessary.

We considered external data as a sample which is not observed or known at the time when the trial is designed. In real-life settings by the nature historical data external to the study, the external data may be available at the time the study is being designed and may potentially affect design decisions. However, importantly, availability of historical data does not inform the study design on the magnitude of $\delta$ because the current data are not yet collected. Thus, the critical assumptions on the magnitude of $\delta$ can still be made in a data-independent manner.

Dynamic information borrowing comes with nontrivial complications. Data generating mechanisms depend on both, the parameter of interest and the conflict between the current and external data. This dependence continues to be present in large samples 
of the local asymptotic framework. Consequently, to determine an unambiguous optimality criterion, statistical inference needs additional specifications on the conflict. For estimation tasks, specification a prior distribution on the conflict, implicitly or explicitly, determines an optimality criterion. In our work, we considered integrated mean squared error, but a prior distribution on the conflict determines an unambiguous optimality criterion for other loss functions as well. For hypothesis testing tasks, additional assumptions on the conflict should also be based on data independent considerations. Previously proposed methods for dynamic information borrowing work well for some ranges of the conflict and poorly at others. Thus, data independent choice of a dynamic information borrowing method has to be closely aligned with non-sampling prior knowledge on the magnitude of the conflict. 

Our findings are generally in line with the skepticism expressed in \cite{galwey2017supplementation} on dynamic information borrowing methods. However, under certain implicit or explicit assumptions on the magnitude of the conflict, as our illustrative example showed, dynamic information borrowing can substantially improve statistical inference. 
 

There is no ``free lunch'' for statisticians when they try to dynamically incorporate external information to improve statistical inference (estimation or hypothesis testing tasks). Asymptotic distributions of statistics often become complex and differ from normal. Implicit or explicit data-independent assumptions on the magnitude of the conflict between the current and external data have to made. Our work suggests generalizations of already known dynamic information borrowing estimators and shows how such data-independent assumptions can be imposed to secure a desired compromise between costs and benefits of external information use.  

\section*{Acknowledgment}

This project was partially supported by the Health Resources and Services Administration (HRSA) of the U.S. Department of Health and Human Services (HHS) under R40MC41748 the Maternal and Child Health Secondary Data Analysis Research Program. This information or content and conclusions are those of the author(s) and should not be construed as the official position or policy of, nor should any endorsements be inferred by HRSA, HHS or the U.S. Government.

\section*{Appendix A: Large Sample Properties}

\textbf{Proof of Theorem 1:} Since $\beta = \theta + \delta$ and $\theta  = \beta - \delta = \beta - h/\sqrt{n}$, 
$\sqrt{n}\left(\hat\beta - \theta\right) = \sqrt{\frac{n}{m}}\sqrt{m}\left(\hat\beta - \beta+h/\sqrt{n}\right)$ converges in distribution to $\sqrt{\frac{p}{1-p}}\zeta_2 + h = N\left(h, \frac{p}{1-p}\right)$.

\noindent \textbf{Proof of Theorem 2:} 
 $\sqrt{n}\left(\hat\theta^p - \theta\right) = \frac{n}{n+m}\zeta_1 + \frac{m}{n+m} \sqrt{n} \left(\hat\beta - \beta + h/\sqrt{n}\right) = \frac{n}{n+m}\zeta_1 + \frac{m}{n+m} \left(\sqrt{\frac{n}{m}}\sqrt{m} \left(\hat\beta - \beta \right)+h\right)$ converges in distribution to $p\zeta_1 + (1-p) \left(\sqrt{\frac{p}{1-p}}\zeta_2+ h \right) = 
     p\zeta_1+\sqrt{p(1-p)}\zeta_2 + (1-p)h=:\zeta_p$. 
Note that the asymptotic distribution of 
$\sqrt{n}\left(\hat\theta^p - \theta\right)$ at $\delta=h/\sqrt{n}$ is a linear combination $\zeta_1\sim N(0,1)$ and $\zeta_2\sim N(0,1)$. Then, $\zeta_p \sim N\left((1-p)h,p\right)$.

\noindent \textbf{Proof of Theorem 3:} Since $\hat\xi = \frac{\hat\beta - \hat\theta}{\sqrt{1/n+1/m}} = \frac{\theta  + h/\sqrt{n} + \zeta_2/\sqrt{m} -\theta - \zeta_1/\sqrt{n}}{\sqrt{1/n+1/m}} = \frac{\sqrt{m}(h-\zeta_1) + \sqrt{n}\zeta_2}{\sqrt{n+m}} \to \xi$ and, using theorems \ref{th1} and \ref{th2}, we find 
$$\sqrt{n}\left(\hat\theta^{ttp}_n-\theta\right) \overset{d}{\to} \text{Pr}(\xi^2>c) \zeta_1 + \text{Pr}(\xi^2\le c) \left(p\zeta_1+\sqrt{p(1-p)}\zeta_2 + (1-p)h\right).$$

\noindent \textbf{Proof of Theorem \ref{th4}:} 
At $\delta=h/\sqrt{n}$, $\sqrt{n} \hat\delta \overset{d}{\to} \sqrt{1/(1-p)}\xi = \sqrt{p/(1-p)} \zeta_2 + h - \zeta_1$, which is a non-degenerate random variable. Then, if $\tau \in (0,0.5)$, $n^{-\tau} \vert\hat\delta\vert \overset{p}\to 0$, which makes the penalty term in the ALASSO criterion diverge to infinity. Consequently, ALASSO estimator of $\delta$ converges to zero and $\hat\theta^{alasso}$ converges to the pooled estimator $\hat\theta^{p}$.  One can easily see that if $\tau \in (0,0.5)$, $\hat\theta^{DIB}\left(g,n^{-2\tau}\right)$ also converges to $\hat\theta^p$.

\noindent \textbf{Proof of Theorem \ref{th5}:} 
$\sqrt{n}\left(\theta^{GDIB} (g,s)-\theta\right) \overset{d}{=} \sqrt{n}\left(\hat\theta-\theta\right) + g(sn\hat\delta^2) \sqrt{n}\hat\delta
\overset{d}{\to} \zeta_1 + g\left(\frac{s\xi^2}{1-p}\right)\xi$,
where $\sqrt{n}\hat\delta \overset{d}{=} \sqrt{n\left(n^{-1}+m^{-1}\right)}\hat\xi \overset{d}{\to} \sqrt{\frac{1}{1-p}} \xi$.

\noindent \textbf{Proof of Corollary \ref{cor1}:} Since $\sqrt{n}\hat\delta \overset{d}{\to}  \xi/\sqrt{1-p}$,
$$\hat{\gamma}^{eb} = \frac{n/m}{\max\{n\hat\delta^2,1+n/m\}-1} \overset{d}{\to} \frac{p/(1-p)}{\max\{\xi^2/(1-p),1/(1-p)\}-1} = \gamma^{eb}$$ and
$\hat{\gamma}^{hd} = \left(1-\sqrt{1-\exp\left(-n\hat\delta^2/8\right)}\right)^2 \overset{d}{\to} \left(1-\sqrt{1-\exp\left(-\xi^2/(8-8p)\right)}\right)^2 = \gamma^{hd}$. Then,
$$\sqrt{n}\left(\hat\theta^{hd}-\theta\right) \overset{d}{=} \sqrt{n}\left(\hat\theta-\theta\right) + \frac{1}{1 +n/(m\hat\gamma^{hd})}\sqrt{n}\hat\delta
\overset{d}{\to} \zeta_1 + \frac{1-p}{1-p+p/\gamma^{hd}}\xi,$$
and, similarly, $\sqrt{n}\left(\hat\theta^{eb}-\theta\right) 
\overset{d}{\to} \zeta_1 + \frac{1-p}{1-p+p/\gamma^{eb}}\xi$,
where $\sqrt{n}\hat\delta \overset{d}{=} \sqrt{n\left(n^{-1}+m^{-1}\right)}\hat\xi \overset{d}{\to} \sqrt{\frac{1}{1-p}} \xi$.

\noindent \textbf{Proof of Corollary \ref{cor2}:} 
$\sqrt{n}\left(\hat\theta^{ammse}-\theta\right) \overset{d}{=} \sqrt{n}\left(\hat\theta-\theta\right) + \frac{1}{n/m +1 +n\hat\delta^2} \sqrt{n}\hat\delta
\overset{d}{\to} \zeta_1 + \frac{\sqrt{1-p}}{1+\xi^2}\xi$,
where $\sqrt{n}\hat\delta \overset{d}{=} \sqrt{n\left(n^{-1}+m^{-1}\right)}\hat\xi \overset{d}{\to} \sqrt{\frac{1}{1-p}} \xi$. 

\noindent \textbf{Proof of Theorem \ref{th6}:} Proof directly follows from Theorem \ref{th4}. 
The asymptotic distribution of $\sqrt{n}\left(\hat\theta_n^{alasso}(\tau)-\theta\right)$ and $\sqrt{n}\left(\hat\theta_n^{DIB}(\tau)-\theta\right)$ is the same of the distribution of the pooled estimator: $N((1-p)h,p)$ if $\delta=h/\sqrt{n}$ including $\delta=0$.  Then, the critical value to secure $\alpha$ for testing $h_{\theta}=0$ versus $h_{\theta}>0$ is $(1-p)h + \sqrt{p}z_{1-\alpha}$. If a researcher wants to control T1ER $\forall h$, a maximum critical value should be chosen. As $h$ increases the critical value, which is a linear function of $h$ also increases and the maximum critical value = $+\infty$. Then, it is not possible to control T1ER at a predetermined level and asymptotic power to detect $h_{\theta}$ is zero.

\section*{Appendix B}
\label{APP:simulations} 

\begin{figure*}[t]
   \centering   
	\includegraphics[width=5.8in]{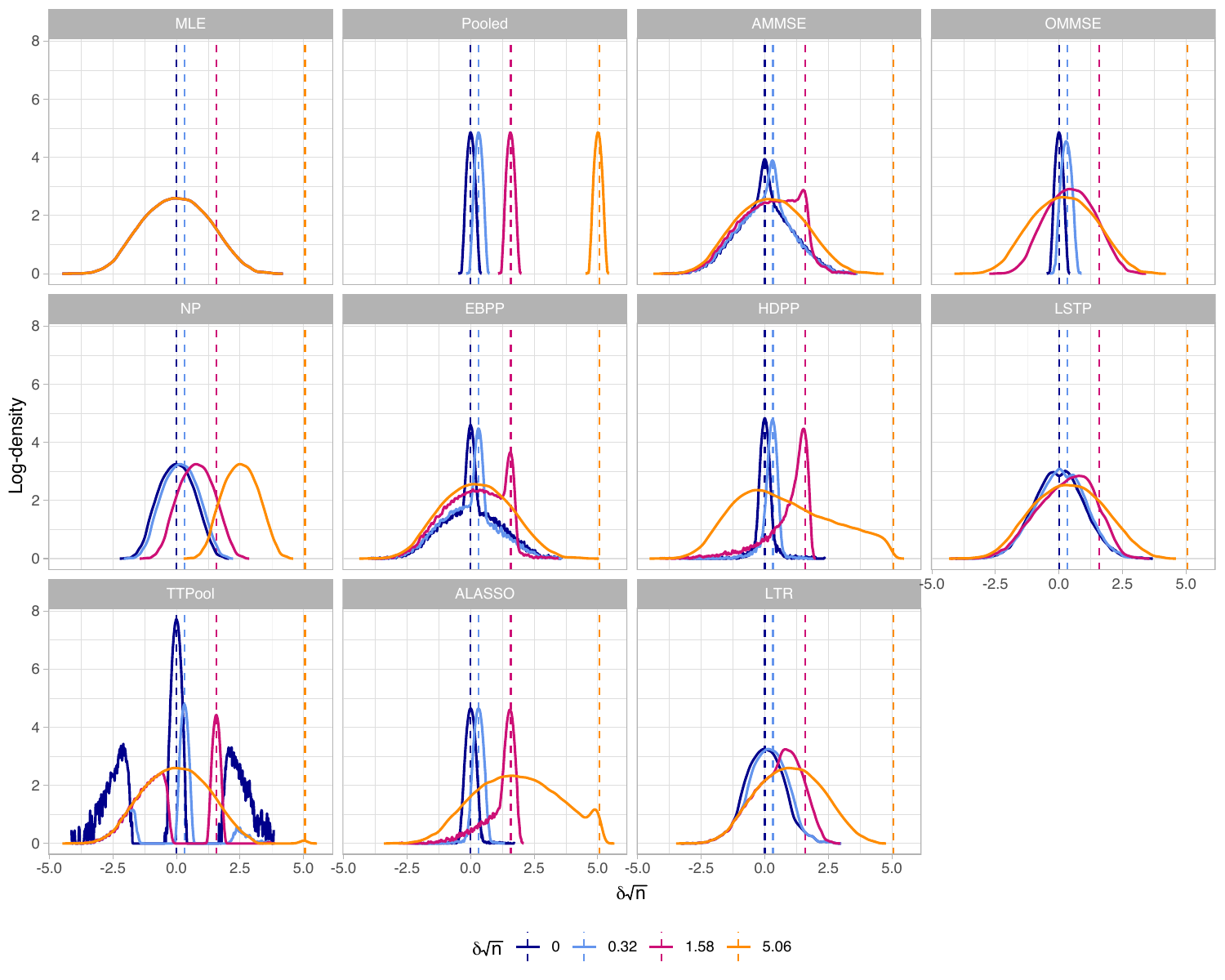}
    \caption{\label{fig:9est} Distributions (log-densities) of estimators of $\theta=0$ under four conflict values: $\delta\in\{0,0.32,1.58,5.06\}$; n=$10^3$, m=$10^5$, and $5\times10^4$ Monte Carlo samples.}
\end{figure*}

Figure \ref{fig:9est} shows DIB densities $\sqrt{n}\delta \in \{0,0.32, 1.58, 5.06\}$, $n=1,000$ and $m=100,000$. 

\bibliographystyle{unsrtnat} 
\bibliography{bib.bib}       

\begin{thebibliography}{47}
\providecommand{\natexlab}[1]{#1}
\providecommand{\url}[1]{\texttt{#1}}
\expandafter\ifx\csname urlstyle\endcsname\relax
  \providecommand{\doi}[1]{doi: #1}\else
  \providecommand{\doi}{doi: \begingroup \urlstyle{rm}\Url}\fi

\bibitem[Pocock and Simon(1975)]{pocock1975}
Stuart~J Pocock and Richard Simon.
\newblock Sequential treatment assignment with balancing for prognostic factors
  in the controlled clinical trial.
\newblock \emph{Biometrics}, pages 103--115, 1975.

\bibitem[Freidlin and Korn(2013)]{freidlin2013borrowing}
Boris Freidlin and Edward~L Korn.
\newblock Borrowing information across subgroups in phase ii trials: is it
  useful?
\newblock \emph{Clinical Cancer Research}, 19\penalty0 (6):\penalty0
  1326--1334, 2013.

\bibitem[Jin et~al.(2023)Jin, Kim, Scheffler, and Jiang]{jin2023bayesian}
Huaqing Jin, Mi-Ok Kim, Aaron Scheffler, and Fei Jiang.
\newblock Bayesian adaptive design for covariate-adaptive historical control
  information borrowing.
\newblock \emph{Statistics in Medicine}, 42\penalty0 (29):\penalty0 5338--5352,
  2023.

\bibitem[Galwey(2017)]{galwey2017supplementation}
NW~Galwey.
\newblock Supplementation of a clinical trial by historical control data: is
  the prospect of dynamic borrowing an illusion?
\newblock \emph{Statistics in Medicine}, 36\penalty0 (6):\penalty0 899--916,
  2017.

\bibitem[Viele et~al.(2014)Viele, Berry, Neuenschwander, Amzal, Chen, Enas,
  Hobbs, Ibrahim, Kinnersley, Lindborg, et~al.]{viele2014use}
Kert Viele, Scott Berry, Beat Neuenschwander, Billy Amzal, Fang Chen, Nathan
  Enas, Brian Hobbs, Joseph~G Ibrahim, Nelson Kinnersley, Stacy Lindborg,
  et~al.
\newblock Use of historical control data for assessing treatment effects in
  clinical trials.
\newblock \emph{Pharmaceutical statistics}, 13\penalty0 (1):\penalty0 41--54,
  2014.

\bibitem[Tarima and Dmitriev(2009)]{Tarima2009}
SS~Tarima and YG~Dmitriev.
\newblock Statistical estimation with possibly incorrect model assumptions.
\newblock \emph{Bul. Tomsk St. University: cont., comput., inf.}, 8:\penalty0
  78--99, 2009.

\bibitem[Dmitriev and Tarassenko(2015)]{yu2015adaptive}
Yury Dmitriev and Peter Tarassenko.
\newblock On adaptive estimation using a prior guess.
\newblock In \emph{Proceedings The International Workshop, Applied Methods of
  Statistical Analysis. Nonparametric Approach}, pages 14--15. Novosibirsk,
  Russia. Novosibirsk, 2015.

\bibitem[Zou(2006)]{zou2006adaptive}
Hui Zou.
\newblock The adaptive lasso and its oracle properties.
\newblock \emph{Journal of the American statistical association}, 101\penalty0
  (476):\penalty0 1418--1429, 2006.

\bibitem[Li et~al.(2023)Li, Lin, Huang, Tian, and Zhu]{li2023frequentist}
Ruilin Li, Ray Lin, Jiangeng Huang, Lu~Tian, and Jiawen Zhu.
\newblock A frequentist approach to dynamic borrowing.
\newblock \emph{Biometrical Journal}, page 2100406, 2023.

\bibitem[Ibrahim and Chen(2000)]{ibrahim2000}
Joseph~G. Ibrahim and Ming-Hui Chen.
\newblock {Power Prior Distributions for Regression Models}.
\newblock \emph{Statistical Science}, 15\penalty0 (1):\penalty0 46--60, 2000.
\newblock ISSN 08834237.
\newblock URL \url{http://www.jstor.org/stable/2676676}.

\bibitem[Ibrahim et~al.(2015)Ibrahim, Chen, Gwon, and Chen]{ibrahim2015}
Joseph Ibrahim, Ming~Hui Chen, Yeongjin Gwon, and Fang Chen.
\newblock {The Power Prior: Theory and Applications}.
\newblock \emph{Statistics in Medicine}, 34\penalty0 (28):\penalty0 3724--3749,
  09 2015.
\newblock \doi{10.1002/sim.6728}.

\bibitem[Gravestock et~al.(2017)Gravestock, Held, and
  consortium]{gravestock2017adaptive}
Isaac Gravestock, Leonhard Held, and COMBACTE-Net consortium.
\newblock Adaptive power priors with empirical bayes for clinical trials.
\newblock \emph{Pharmaceutical statistics}, 16\penalty0 (5):\penalty0 349--360,
  2017.

\bibitem[Calderazzo et~al.(2023)Calderazzo, Tarima, Reid, Flournoy, Friede,
  Geller, Rosenberger, Stallard, Ursino, Vandemeulebroecke,
  et~al.]{calderazzo2023coping}
Silvia Calderazzo, Sergey Tarima, Carissa Reid, Nancy Flournoy, Tim Friede,
  Nancy Geller, James~L Rosenberger, Nigel Stallard, Moreno Ursino, Marc
  Vandemeulebroecke, et~al.
\newblock Coping with information loss and the use of auxiliary sources of
  data: A report from the niss ingram olkin forum series on unplanned clinical
  trial disruptions.
\newblock \emph{Statistics in Biopharmaceutical Research}, pages 1--17, 2023.

\bibitem[Efron and Morris(1972)]{efron1972limiting}
Bradley Efron and Carl Morris.
\newblock Limiting the risk of bayes and empirical bayes estimators—part ii:
  The empirical bayes case.
\newblock \emph{Journal of the American Statistical Association}, 67\penalty0
  (337):\penalty0 130--139, 1972.

\bibitem[Qin et~al.(2022)Qin, Liu, and Li]{qin2022selective}
Jing Qin, Yukun Liu, and Pengfei Li.
\newblock A selective review of statistical methods using calibration
  information from similar studies.
\newblock \emph{Statistical Theory and Related Fields}, pages 1--16, 2022.

\bibitem[Tarima and Pavlov(2006)]{tarima2006}
S~Tarima and D~Pavlov.
\newblock Using auxiliary information in statistical function estimation.
\newblock \emph{ESAIM: Probab. Stat.}, 10:\penalty0 11--23, 2006.

\bibitem[Chen et~al.(2024)Chen, Han, Chen, Shardell, and
  Qin]{chen2024integrating}
Chixiang Chen, Peisong Han, Shuo Chen, Michelle Shardell, and Jing Qin.
\newblock Integrating external summary information in the presence of prior
  probability shift: an application to assessing essential hypertension.
\newblock \emph{Biometrics}, 80\penalty0 (3):\penalty0 ujae090, 2024.

\bibitem[Ferguson(2017)]{ferguson2017course}
Thomas~S Ferguson.
\newblock \emph{A course in large sample theory}.
\newblock Routledge, 2017.

\bibitem[Nikitin(1995)]{nikitin1995asymptotic}
Yakov~Yurievich Nikitin.
\newblock \emph{Asymptotic efficiency of nonparametric tests}.
\newblock Cambridge University Press, 1995.

\bibitem[Cam(1960)]{cam1960locally}
Le~Cam.
\newblock Locally asymptotically normal families of distributions. certain
  approximations to families of distributions and their use in the theory of
  estimation and testing hypotheses.
\newblock \emph{Univ. California Publ. Statist.}, 3:\penalty0 37, 1960.

\bibitem[{Van der Vaart}(1998)]{vandervaart1998}
A.W. {Van der Vaart}.
\newblock \emph{Asymptotic Statistics}.
\newblock Cambridge University Press, 1998.

\bibitem[Sidak et~al.(1999)Sidak, Sen, and Hajek]{sidak1999theory}
Zbynek Sidak, Pranab~K Sen, and Jaroslav Hajek.
\newblock \emph{Theory of rank tests}.
\newblock Elsevier, 1999.

\bibitem[{Le Cam}(1960)]{lecam1960}
Lucien~M. {Le Cam}.
\newblock Locally asymptotically normal families of distributions.
\newblock \emph{Univ. California Publ. Statist.}, 3:\penalty0 37--98, 1960.

\bibitem[Kahn and Gu{\c{t}}{\u{a}}(2009)]{kahn2009local}
Jonas Kahn and M{\u{a}}d{\u{a}}lin Gu{\c{t}}{\u{a}}.
\newblock Local asymptotic normality for finite dimensional quantum systems.
\newblock \emph{Communications in Mathematical Physics}, 289\penalty0
  (2):\penalty0 597--652, 2009.

\bibitem[Tarima and Flournoy(2024)]{tarima2024cost}
Sergey Tarima and Nancy Flournoy.
\newblock The cost of sequential adaptation and the lower bound for mean
  squared error.
\newblock \emph{Statistical Papers}, pages 1--25, 2024.

\bibitem[Tarima et~al.(2020)Tarima, Tuyishimire, Sparapani, Rein, and
  Meurer]{tarima2020estimation}
Sergey Tarima, Bonifride Tuyishimire, Rodney Sparapani, Lisa Rein, and John
  Meurer.
\newblock Estimation combining unbiased and possibly biased estimators.
\newblock \emph{Journal of Statistical Theory and Practice}, 14\penalty0
  (2):\penalty0 1--20, 2020.

\bibitem[Tibshirani(1996)]{tibshirani1996regression}
Robert Tibshirani.
\newblock Regression shrinkage and selection via the lasso.
\newblock \emph{Journal of the Royal Statistical Society Series B: Statistical
  Methodology}, 58\penalty0 (1):\penalty0 267--288, 1996.

\bibitem[Kanapka and Ivanova(2024)]{kanapka2024frequentist}
Lauren Kanapka and Anastasia Ivanova.
\newblock A frequentist design for basket trials using adaptive lasso.
\newblock \emph{Statistics in Medicine}, 43\penalty0 (1):\penalty0 156--172,
  2024.

\bibitem[Neuenschwander et~al.(2010)Neuenschwander, Capkun-Niggli, Branson, and
  Spiegelhalter]{neuenschwander2010}
Beat Neuenschwander, Gorana Capkun-Niggli, Michael Branson, and David~J
  Spiegelhalter.
\newblock Summarizing historical information on controls in clinical trials.
\newblock \emph{Clinical Trials}, 7\penalty0 (1):\penalty0 5--18, 2010.
\newblock \doi{10.1177/1740774509356002}.
\newblock URL \url{https://doi.org/10.1177/1740774509356002}.
\newblock PMID: 20156954.

\bibitem[Schmidli et~al.(2014)Schmidli, Gsteiger, Roychoudhury, O'Hagan,
  Spiegelhalter, and Neuenschwander]{schmidli2014}
Heinz Schmidli, Sandro Gsteiger, Satrajit Roychoudhury, Anthony O'Hagan, David
  Spiegelhalter, and Beat Neuenschwander.
\newblock Robust meta-analytic-predictive priors in clinical trials with
  historical control information.
\newblock \emph{Biometrics}, 70\penalty0 (4):\penalty0 1023--1032, 2014.
\newblock ISSN 0006341X, 15410420.
\newblock URL \url{http://www.jstor.org/stable/24538386}.

\bibitem[Hobbs et~al.(2012)Hobbs, Sargent, and Carlin]{hobbs2012}
Brian~P. Hobbs, Daniel~J. Sargent, and Bradley~P. Carlin.
\newblock Commensurate priors for incorporating historical information in
  clinical trials using general and generalized linear models.
\newblock \emph{Bayesian Analysis}, 7\penalty0 (3):\penalty0 639 -- 674, 2012.
\newblock \doi{10.1214/12-BA722}.
\newblock URL \url{https://doi.org/10.1214/12-BA722}.

\bibitem[Neuenschwander and Schmidli(2020)]{neuenschwander2020}
Beat Neuenschwander and Heinz Schmidli.
\newblock Use of historical data.
\newblock In \emph{Bayesian Methods in Pharmaceutical Research}, pages
  111--137. Chapman and Hall/CRC, 2020.

\bibitem[Wiesenfarth and Calderazzo(2020)]{wiesenfarth2019}
Manuel Wiesenfarth and Silvia Calderazzo.
\newblock Quantification of prior impact in terms of effective current sample
  size.
\newblock \emph{Biometrics}, 76\penalty0 (1):\penalty0 326--336, 2020.
\newblock \doi{https://doi.org/10.1111/biom.13124}.
\newblock URL \url{https://onlinelibrary.wiley.com/doi/abs/10.1111/biom.13124}.

\bibitem[Pawel et~al.(2024)Pawel, Aust, Held, and Wagenmakers]{pawel2022}
Samuel Pawel, Frederik Aust, Leonhard Held, and Eric-Jan Wagenmakers.
\newblock Power priors for replication studies.
\newblock \emph{TEST}, 33\penalty0 (1):\penalty0 127--154, 2024.

\bibitem[Thompson et~al.(2021)Thompson, Chu, Xu, Li, Nair, and
  Tiwari]{thompson2021dynamic}
Laura Thompson, Jianxiong Chu, Jianjin Xu, Xuefeng Li, Rajesh Nair, and Ram
  Tiwari.
\newblock Dynamic borrowing from a single prior data source using the
  conditional power prior.
\newblock \emph{Journal of Biopharmaceutical Statistics}, 31\penalty0
  (4):\penalty0 403--424, 2021.

\bibitem[Nikolakopoulos et~al.(2018)Nikolakopoulos, van~der Tweel, and
  Roes]{nikolakopoulos2018dynamic}
Stavros Nikolakopoulos, Ingeborg van~der Tweel, and Kit~CB Roes.
\newblock Dynamic borrowing through empirical power priors that control type i
  error.
\newblock \emph{Biometrics}, 74\penalty0 (3):\penalty0 874--880, 2018.

\bibitem[Ollier et~al.(2020)Ollier, Morita, Ursino, and Zohar]{ollier2020}
Adrien Ollier, Satoshi Morita, Moreno Ursino, and Sarah Zohar.
\newblock An adaptive power prior for sequential clinical trials--application
  to bridging studies.
\newblock \emph{Statistical methods in medical research}, 29\penalty0
  (8):\penalty0 2282--2294, 2020.

\bibitem[Dawid(1973)]{dawid1973}
A.~P. Dawid.
\newblock {Posterior expectations for large observations}.
\newblock \emph{Biometrika}, 60\penalty0 (3):\penalty0 664--667, 12 1973.
\newblock ISSN 0006-3444.
\newblock \doi{10.1093/biomet/60.3.664}.
\newblock URL \url{https://doi.org/10.1093/biomet/60.3.664}.

\bibitem[O'Hagan(1979)]{ohagan1979}
Anthony O'Hagan.
\newblock On outlier rejection phenomena in bayes inference.
\newblock \emph{Journal of the Royal Statistical Society. Series B
  (Methodological)}, 41\penalty0 (3):\penalty0 358--367, 1979.
\newblock ISSN 00359246.
\newblock URL \url{http://www.jstor.org/stable/2985064}.

\bibitem[Tarima et~al.(2013)Tarima, Vexler, and Singh]{Tarima2013}
SS~Tarima, A~Vexler, and S~Singh.
\newblock Robust mean estimation under a possibly incorrect log-normality
  assumption.
\newblock \emph{Commun. Stat.--Simul. C.}, 42\penalty0 (2):\penalty0 316--326,
  2013.

\bibitem[Van~Lancker et~al.(2023)Van~Lancker, Tarima, Bartlett, Bauer,
  Bharani-Dharan, Bretz, Flournoy, Michiels, Parra, Rosenberger,
  et~al.]{van2023rejoinder}
Kelly Van~Lancker, Sergey Tarima, Jonathan Bartlett, Madeline Bauer, Bharani
  Bharani-Dharan, Frank Bretz, Nancy Flournoy, Hege Michiels, Camila~Olarte
  Parra, James~L Rosenberger, et~al.
\newblock Rejoinder: Estimands and their estimators for clinical trials
  impacted by the covid-19 pandemic: A report from the niss ingram olkin forum
  series on unplanned clinical trial disruptions.
\newblock \emph{Statistics in Biopharmaceutical Research}, 15\penalty0
  (1):\penalty0 119--124, 2023.

\bibitem[DeGroot(2005)]{degroot2005optimal}
Morris~H DeGroot.
\newblock \emph{Optimal statistical decisions}.
\newblock John Wiley \& Sons, 2005.

\bibitem[Parmigiani and Inoue(2009)]{parm2009}
Giovanni Parmigiani and Lurdes Inoue.
\newblock \emph{Decision theory: principles and approaches}, volume 812.
\newblock John Wiley \& Sons, 2009.

\bibitem[Kopp-Schneider et~al.(2020)Kopp-Schneider, Calderazzo, and
  Wiesenfarth]{kopp2020power}
Annette Kopp-Schneider, Silvia Calderazzo, and Manuel Wiesenfarth.
\newblock Power gains by using external information in clinical trials are
  typically not possible when requiring strict type i error control.
\newblock \emph{Biometrical Journal}, 62\penalty0 (2):\penalty0 361--374, 2020.

\bibitem[Kopp-Schneider et~al.(2023)Kopp-Schneider, Wiesenfarth, Held, and
  Calderazzo]{kopp2024}
Annette Kopp-Schneider, Manuel Wiesenfarth, Leonhard Held, and Silvia
  Calderazzo.
\newblock Simulating and reporting frequentist operating characteristics of
  clinical trials that borrow external information: Towards a fair comparison
  in case of one-arm and hybrid control two-arm trials.
\newblock \emph{Pharmaceutical Statistics}, 2023.
\newblock \doi{https://doi.org/10.1002/pst.2334}.
\newblock URL \url{https://onlinelibrary.wiley.com/doi/abs/10.1002/pst.2334}.

\bibitem[Shulman et~al.(2018)Shulman, D’Angelo, Harrison, Smith, and
  Warner]{shulman_pregnancy_2018}
Holly~B. Shulman, Denise~V. D’Angelo, Leslie Harrison, Ruben~A. Smith, and
  Lee Warner.
\newblock The {Pregnancy} {Risk} {Assessment} {Monitoring} {System} ({PRAMS}):
  {Overview} of {Design} and {Methodology}.
\newblock \emph{American Journal of Public Health}, 108\penalty0 (10):\penalty0
  1305--1313, August 2018.
\newblock ISSN 0090-0036.
\newblock \doi{10.2105/AJPH.2018.304563}.
\newblock URL
  \url{https://ajph.aphapublications.org/doi/abs/10.2105/AJPH.2018.304563}.

\bibitem[Pericchi and Smith(1992)]{peri1992}
L.~R. Pericchi and A.~F.~M. Smith.
\newblock Exact and approximate posterior moments for a normal location
  parameter.
\newblock \emph{Journal of the Royal Statistical Society. Series B
  (Methodological)}, 54\penalty0 (3):\penalty0 793--804, 1992.
\newblock ISSN 00359246.
\newblock URL \url{http://www.jstor.org/stable/2345859}.

\end{thebibliography}

\end{document}